\documentclass[useAMS,usenatbib]{mn2e}

\usepackage{amsmath} \usepackage{amssymb} \usepackage{aasmacros}
\usepackage[pdftex]{graphicx} \newcommand{\oi}{{\mbox{O\,{\sc i}}} }
\newcommand{\hi}{{\mbox{H\,{\sc i}}} }
\newcommand{\civ}{{\mbox{C\,{\sc iv}}} }
\newcommand{\lya}{{Lyman-$\alpha$~}}

\title[Probing the the circumgalactic medium at $z\ga 6$ with  OI
  absorption]{Probing the metallicity and ionization state  of the
  circumgalactic medium at  $z\sim 6$ and  beyond with  OI absorption}

\author[L. C. Keating et al.] {Laura C.  Keating$^{1,2}$\thanks{E-mail:
    lck35@ast.cam.ac.uk}, Martin G. Haehnelt$^{1,2}$, George
  D. Becker$^{1,2}$ \& James S. Bolton$^{3}$\\ 
$^1$ Institute of Astronomy, University of Cambridge,
  Madingley Road, Cambridge, CB3 0HA\\
$^2$ Kavli Institute  for Cosmology,  University of Cambridge,
  Madingley Road, Cambridge, CB3 0HA\\
$^3$ School of Physics and
  Astronomy, University of Nottingham, University Park, Nottingham,
  NG7 2RD}

\begin{document}

\date{August 2013}

\pagerange{\pageref{firstpage}--\pageref{lastpage}} 

\pubyear{2002}

\maketitle

\label{firstpage}

\begin{abstract}
Low ionization metal absorption due to \oi has been identified as an
important probe of the physical state of the inter-/circumgalactic
medium at the tail-end of reionization. We use here high-resolution
hydrodynamic simulations to interpret the incidence rate of \oi
absorbers at $z \sim 6$ as observed by \citet{becker2011}.  We
  infer  weak \oi absorbers (EW $\ga$ 0.1 \AA) to  have typical \hi
column densities  in the range of sub-DLAs, densities of 80 times the
mean  baryonic density and metallicities of about 1/500 th solar. This
is  similar to the metallicity  inferred at similar overdensities at
$z\sim 3$,  suggesting  that the metal enrichment  of the
circumgalactic medium around low-mass galaxies has already progressed
considerably by $z\sim 6$.  The apparently rapid evolution of the
incidence rates for \oi absorption over the redshift range $5 \lesssim
z \lesssim 6$ mirrors that of self-shielded Lyman-Limit systems at
lower redshift and is mainly due to the rapid decrease of the
meta-galactic photo-ionization rate at $z\ga 5$.  We predict the
incidence rate of \oi absorbers to continue to  rise rapidly with
increasing redshift as the IGM becomes more neutral.  If the
distribution of metals extends to  lower density regions, \oi
absorbers will allow the metal enrichment  of the increasingly neutral
filamentary structures of the cosmic web to be probed.  
\end{abstract}


\section{Introduction}

\lya absorption line studies are an  important tool to study the
ionization state of the IGM at the tail-end of reionization
(e.g. \citealt{songaila2004}; \citealt{fan2006}; \citealt{bolton2007};
\citealt{becker2007}; \citealt{mcquinn2008}; \citealt{mesinger2010};
\citealt{becker2013}).  At $z > 6$, however,   little detailed
information can be obtained from \lya absorption about the spatial
distribution of the neutral gas, as a significant fraction of the IGM
is opaque to  Lyman series photons. As pointed out by \citet{oh2002},
\oi  absorption provides an interesting alternative to study the
distribution of neutral metal enriched gas in the IGM.  \oi has an
atomic transition with a rest wavelength longer than Ly$\alpha$, so is
visible redward of the Ly$\alpha$ emission.  \oi also has an
ionization energy close to that of neutral hydrogen  ($\Delta E =
0.019 \, \rmn{eV}$) and is an excellent tracer of (self-shielded)
neutral gas.

\begin{figure*}
\includegraphics[width=2.2\columnwidth]{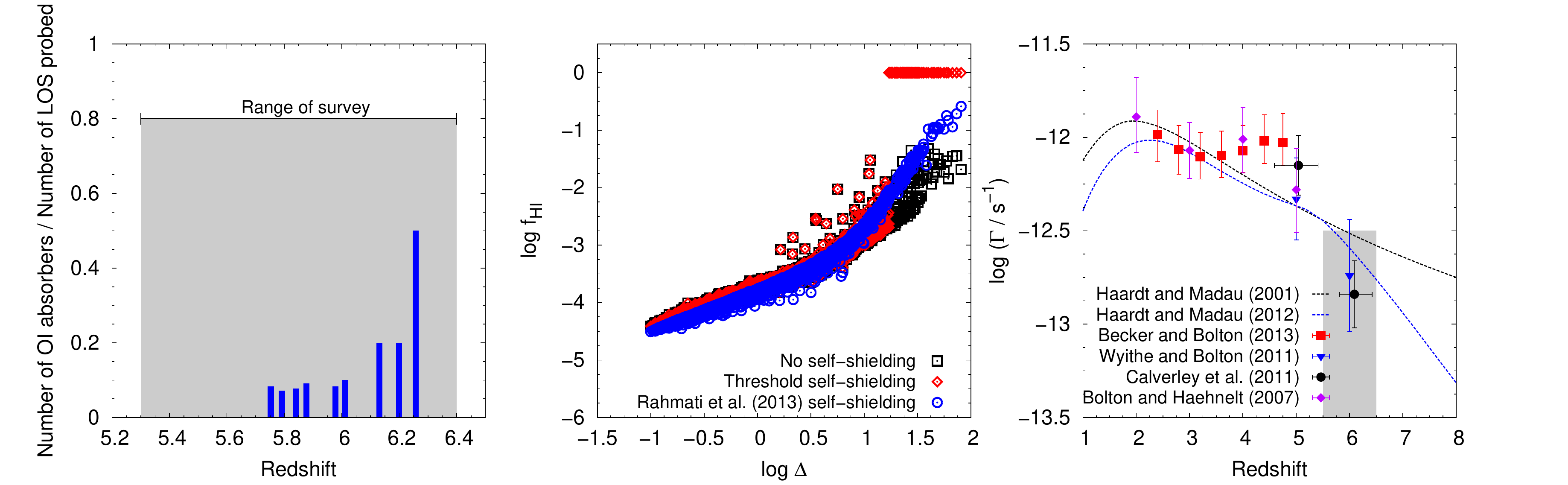}
\caption{Left: The incidence rate of \oi absorbers in the survey  of
  \citet{becker2011}. Nine systems containing \mbox{O\,{\sc i}} were
  detected. The height of the bars is the ratio of the number of \oi
  absorbers to the number of sightlines that were surveyed at the
  redshift of the absorber. All of the \mbox{O\,{\sc i}} absorption
  systems were found at $z > 5.75$ and the line of sight number
  density of absorbers appears to increase rapidly with increasing
  redshift. Middle: The dependence of neutral fraction on  overdensity
  for the simulation with no self-shielding, with the threshold
  self-shielding model used in \citet{bolton2013} and the
  self-shielding prescription used in \citet{rahmati2013}.  The
  photo-ionization rate in this case was taken to be $\log (\Gamma /
  \textnormal{s}^{-1}) = -12.8$. Right: A plot of the background
  photo-ionization rate $\Gamma$ against redshift $z$. The points are
  measurements of the UV background at different redshifts taken from
  \citet{becker2013}, \citet{wyithe2011}, \citet{calverley2011} and
  \citet{bolton2007}. The \citet{haardt2001} and \citet{haardt2012}
  models for $\Gamma$ are also shown. The shaded grey region shows the
  range of $\Gamma$ consistent with  the observed incidence rate of
  \oi absorbers  at $z\sim 6$.} 
  \label{survey_hfrac_gamma}
\end{figure*}

The most comprehensive searches for \oi  absorption  at high redshift
so far have  been performed by \citet{becker2007} and
\citet{becker2011}. The results of the \citet{becker2011} survey are
summarised in the left panel of Figure \ref{survey_hfrac_gamma} and
Table 1. The survey was based on  spectra of  17 QSOs with emission
redshifts ranging from  $5.8 < z_{\textnormal{\scriptsize{em}}} <
6.4$. High-resolution spectra were obtained  for nine of the QSOs,
using Keck (HIRES) or Magellan (MIKE) and moderate-resolution spectra
were obtained for the rest using Keck (ESI). Ten low-ionization metal
systems were detected, with nine of these systems containing
\mbox{O\,{\sc i}} lines. The \mbox{O\,{\sc i}} absorption systems are
summarised in Table 1. The total absorption path-length of the 2011
survey was  $\Delta X = 39.5$, where $X$ is defined as 
\begin{equation}
X(z) = \int^z_0 \, \frac{H_0}{H(z')} (1+z')^2 \, \rmn{d}z',
\end{equation}
and $H(z)$ is the Hubble constant at redshift $z$ \citep{bahcall1969}.

Even though the survey path extends from $z=5.3$,  all ten detected
systems occur at $z > 5.75$  and, as the left panel of Figure
\ref{survey_hfrac_gamma} demonstrates,  the incidence of
low-ionization systems appears to increase rapidly  with increasing
redshift.  \citet{becker2011} suggested that this rapid evolution is
due to   the evolution of the meta-galactic UV background at $z \sim
6$.  They further pointed out that the overall incidence rate is
comparable   to that of Damped \lya systems (DLAs) and Lyman-Limit
systems (LLSs) at $z \sim 3$ and proposed  that the absorbers may be
probing the circumgalactic medium of faint galaxies at high redshift (see \citet{kulkarni2013} and \citet{maio2013} for
  recent  modelling of DLA host galaxies and  metal enrichment  at
  these high redshifts).

\begin{table}
  \caption{Equivalent width of the \mbox{O\,{\sc i}} lines observed by
    \citet{becker2011}}
  \begin{tabular}{@{}lccc@{}}
    \hline QSO & $z_{\textnormal{\scriptsize{abs}}}$ & Instrument &
    $W_{\lambda}$ (\AA)\\ \hline SDSS J2054-0005 & 5.9776 & ESI &
    0.124\\ SDSS J2315-0023 & 5.7529 & ESI & 0.238\\ SDSS J0818+1722 &
    5.7911 & HIRES & 0.182\\ SDSS J0818+1722 & 5.8765 & HIRES &
    0.058\\ SDSS J1623+3112 & 5.8415 & HIRES & 0.391\\ SDSS J1148+5251
    & 6.0115 & HIRES & 0.162\\ SDSS J1148+5251 & 6.1312 & HIRES &
    0.079\\ SDSS J1148+5251 & 6.1988 & HIRES & 0.020\\ SDSS J1148+5251
    & 6.2575 & HIRES & 0.036\\ \hline
  \end{tabular}
\end{table}

There are still many questions to be answered, however, about the
spatial distribution of the gas giving rise to this absorption and the
physical properties of the absorbing gas, as well as the link between
these absorbers and self-shielded absorption systems at lower
redshift. While the incidence rate of DLAs appears to evolve rather
slowly with redshift (e.g. \citealt{seyffert2013}),  the incidence
rate of LLSs is increasing rapidly with increasing redshift
\citep{fumagalli2013, songaila2010}.  \citeauthor{bolton2013} (2013,
BH13) have  recently emphasized  that this rapid evolution of LLSs is
expected to accelerate further as the tail-end of the  epoch of
reionization is probed. 

In this paper, we will use the same hydrodynamical simulation  used in
BH13 to reproduce  the observed  damping wing redward of the \lya
emission in the  $z=7.085$ QSO  ULASJ1120+0641 \citep{mortlock2011} to
model the \oi absorbers  of \citet{becker2011}. The simulations have
been shown to reproduce well the properties of the \lya forest in QSO
absorption spectra over a  wide redshift range  $2\la z \la 6$ and
should allow us to obtain  a reasonable representation of the spatial
distribution of neutral self-shielded  gas at $z\ge 6$, where the
meta-galactic photo-ionization rate is expected to drop rapidly. To
model the \oi absorption, we adopt a relatively simple approach, where
we assume a simple  power-law metallicity-density relation and apply
self-shielding corrections to our simulations, which do not include
radiative transfer. This allows us  to vary the metal distribution 
and ionization state sufficiently to obtain a good match for the data 
and to constrain in this way both the  metal/oxygen distribution and the ionization state
of the CGM/IGM. Recently  \citet{finlator2013} have  
attempted to model the distribution of metals  and self-shielded gas  
from first principles with full  radiative transfer simulations. 
This approach, based on much more  expensive ab initio simulations, 
is offering important complementary insights towards a 
self-consistent picture for the production and transport of metals and
ionizing photons.

We will discuss  the details of our simulations in section 2.  Section
3 will describe our modelling of the \oi absorbers in  the
\citet{becker2011} survey.  In section 4, we will discuss our results
with regard to metallicity and photo-ionization rate of   the
inter-/circumgalactic medium  and present predictions for the
evolution of \oi absorption   at $z>6$. Section 5 gives a summary and
our conclusions. A calculation of the neutral fraction of \oi with the
photo-ionization code CLOUDY \citep{ferland1998} is presented in an appendix.

\section{Modelling the high-redshift IGM}

\begin{figure*}
\includegraphics[width=2.2\columnwidth]{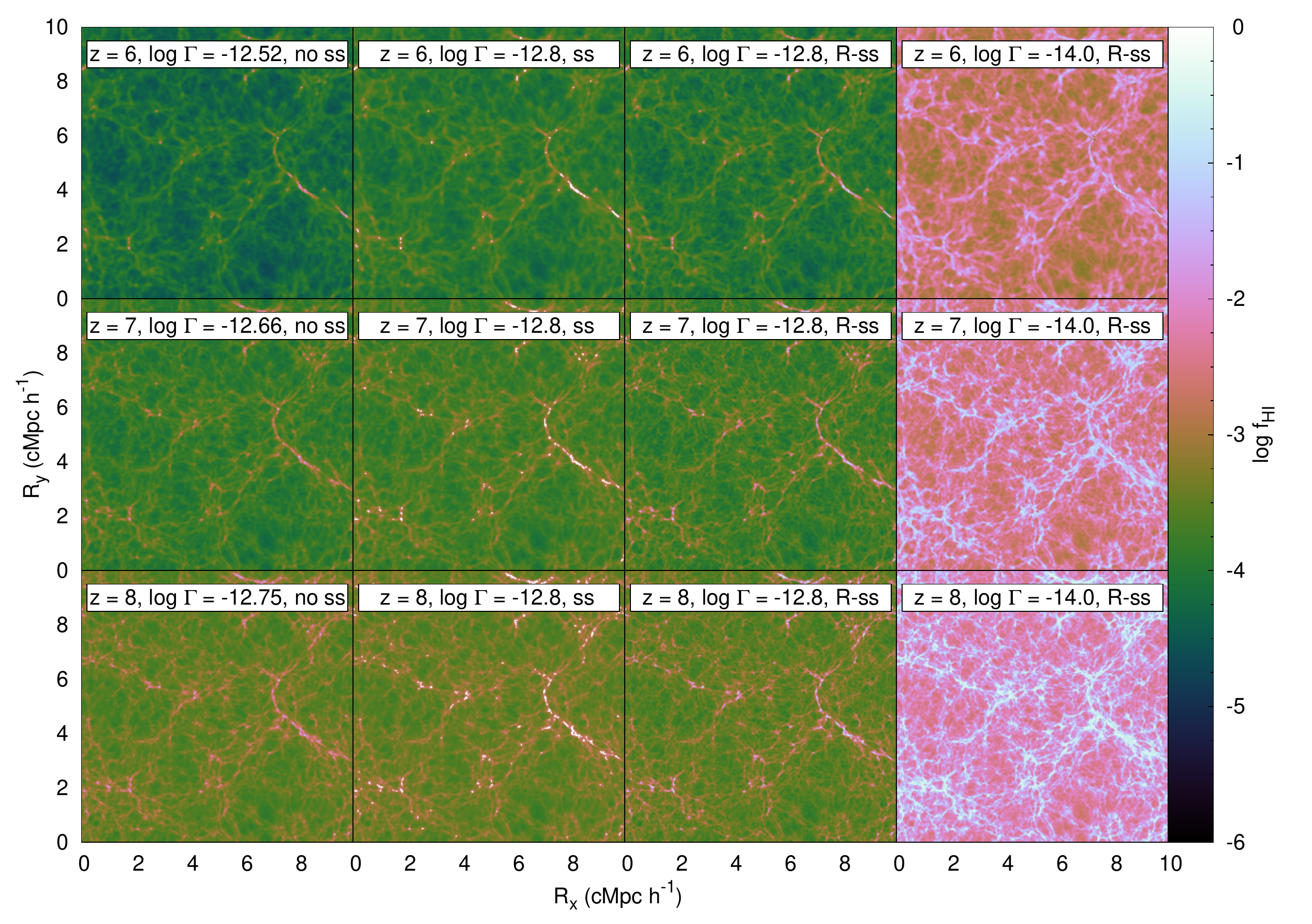}
\caption{The neutral hydrogen fraction of the simulation. The top row
  is at $z=6.0$, the middle row is at $z=7.1$ and the bottom row is at
  $z=8.0$. Shown is a thin slice of thickness  $39h^{-1}
  \,\rmn{ckpc}$ midway through the  simulation box with boxsize
  $10h^{-1}  \,\rmn{cMpc}$. From left:  column 1   does not include
  self-shielding, column 2 shows the threshold self-shielding model
  used in \citet{bolton2013} with $\log (\Gamma / \textnormal{s}^{-1})
  = -12.8$, column 3 includes  self-shielding using the
  \citet{rahmati2013} model with $\log (\Gamma /
  \textnormal{s}^{-1}) = -12.8$, and column 4 includes  self-shielding
  using the \citet{rahmati2013} model with $\log (\Gamma /
  \textnormal{s}^{-1}) = -14.0$.}
  \label{selfshielding}
\end{figure*}

\subsection{Hydrodynamical simulation of the CGM and IGM}

Our modelling is based on a cosmological hydrodynamical simulation
with outputs at redshifts  $z = (6.0, 7.1, 8.0)$. The simulation,
described in detail in BH13, was performed using the  parallel TreeSPH
code \textsc{gadget-3} (the previous version of the code,
\textsc{gadget-2}, is described in \citet{springel2005}). The
simulation has a gas particle mass of $9.2 \times 10^4 h^{-1} \,
\text{M}_{\odot}$ and  a box of size $10 h^{-1} \, \text{cMpc}$ (where
cMpc refers to comoving Mpc). The gravitational softening length was
$0.65 h^{-1} \, \text{ckpc}$. The following values were assumed for
the cosmological parameters  $(h, \Omega_{\text{m}}, \Omega_{\Lambda},
\Omega_{\text{b}} \, h^2, \sigma_8, n) = (0.72, 0.26, 0.74, 0.023,
0.80, 0.96)$.

In the simulation, the IGM  is reionized instantaneously by a uniform
photo-ionizing background at $z=9$. The photo-ionizing background is
based on the \citet{haardt2001} model for emission from quasars and
galaxies. The simulation does not include radiative transfer and  was
performed assuming the optically thin limit for ionizing
radiation. The left column of panels in Figure \ref{selfshielding}
show the neutral  fraction for a thin slice of the simulation midway
through the simulation box in the optically thin limit. The hydrogen
photo ionization rates $\Gamma$ are  those of the \citet{haardt2001}
model as indicated on the plot.  The neutral fraction of the gas
increases with increasing redshift, as expected. No fully neutral
regions (shaded  white) are seen in these optically thin simulations.
Regions surrounded by  neutral hydrogen column densities  with $\log
N_{\textnormal{\scriptsize{HI}}} \ga 17$  are, however,   optically
thick for ionizing radiation and  the gas self-shields.  This  is
neglected in the optically thin approximation. We model this by
post-processing the simulations  with simple models for the
self-shielding as described in the next section.

\subsection{Modelling self-shielded regions}

We used two methods to add self-shielded regions to our simulation in
post-processing. The first method approximates the effect of
self-shielding by assuming a simple density threshold above  which the
gas becomes fully neutral \citep{haehnelt1998}. This is motivated by
numerical simulations of the IGM, which show that the gas  responsible
for the intervening \lya absorption   exhibits a reasonably  tight
correlation between (absorption-weighted) density and  the column
density of the absorbers.  As in BH13 we first applied  a simple
self-shielding model that  assumes  that the absorption length
corresponds to the Jeans length  of gas in photo-ionization
equilibrium as proposed by  \citet{schaye2001}. 

The assumed density threshold is given by 
\begin{equation}
\Delta_{\textnormal{\scriptsize{ss}}} = 36 \, \Gamma^{2/3}_{-12} \,
T^{2/15}_{4} \, \left(\frac{\mu}{0.61}\right)^{1/3} \,
\left(\frac{f_{\textnormal{\scriptsize{e}}}}{1.08} \right)^{-2/3} \,
\left(\frac{1+z}{8}\right)^{-3}.
\end{equation} 
Here, $\Delta = \rho \, / \, \overline{\rho}$ is the overdensity,
$\Gamma_{-12} = \Gamma / 10^{-12} \, \textnormal{s}^{-1}$ is the
background photo-ionization rate and $T_{4} = T / 10^{4} \,
\textnormal{K}$, with $T$ the temperature of the gas. As already
mentioned, this approximation assumes the typical size of an absorber
to be the local Jeans length. It also assumes that the column density
at which a LLS becomes self-shielding is $\log N_{\rmn{HI}} =
17.2$. Note that the case-A recombination coefficient for ionized
hydrogen was used  as given in \citet{abel1997}. 

\citet{rahmati2013} have recently demonstrated that the above
threshold self-shielding model, while being a reasonable first-order
approximation, corresponds to a significantly sharper transition than
predicted by  full radiative transfer simulations that take
recombination radiation into account. We therefore also implemented a
simple fitting formulae suggested by \citet{rahmati2013} based on
their  simulations that include radiative transfer.  In the \citet{rahmati2013} self-shielding model, the  photo-ionization
rate is assumed to be a smooth function of the density which can be
approximated as, 
\begin{equation}
\begin{split}
\frac{\Gamma_{\textnormal{\scriptsize{Phot}}}}{\Gamma_{\textnormal{\scriptsize{HI}}}}
= \, 0.98& \,
\bigg[1+\bigg(\frac{n_{\textnormal{\scriptsize{H}}}}{n_{\textnormal{\scriptsize{H,ss}}}}\bigg)^{1.64}\bigg]^{-2.28}
\\ + \, &0.02 \,
\bigg[1+\frac{n_{\textnormal{\scriptsize{H}}}}{n_{\textnormal{\scriptsize{H,ss}}}}\bigg]^{-0.84}.
\end{split}
\end{equation}
$\Gamma_{\textnormal{\scriptsize{Phot}}}$ is the total
photo-ionization rate and $n_{\textnormal{\scriptsize{H,ss}}}$ is the
characteristic number density for self-shielding, which can be related
to $\Delta_{\textnormal{\scriptsize{ss}}}$. The neutral fraction of
the gas can then be calculated using
$\Gamma_{\textnormal{\scriptsize{Phot}}}$,
$n_{\textnormal{\scriptsize{H}}}$ and the temperature of the gas.

A comparison of the neutral fraction as a function of overdensity for
the simple threshold self-shielding model, the \citet{rahmati2013}
model and the case of no self-shielding is shown  in the middle panel
of Figure \ref{survey_hfrac_gamma} for our simulations at $z=6$ with
$\log(\Gamma / \textnormal{s}^{-1}) = -12.8$ (our fiducial  value consistent 
with \lya forest measurements at $z\sim 6$ based on both  the effective optical
depth and the  proximity effect method shown in the right panel).
In the case of no
self-shielding, the gas in the simulation is highly ionized
everywhere.  Even at the highest overdensities, the neutral fraction
of the gas is still only about one percent.  The \citet{rahmati2013}
model predicts a neutral fraction close to that of our optically thin
simulation   at low overdensities  ($\Delta \lesssim 1$), but the
transition to fully neutral gas  is much more gentle than in the
threshold self-shielding model.  At intermediate overdensities
corresponding to sub-DLA column  densities, the predicted neutral
fraction changes smoothly from 1 percent for  LLS column densities to
100 percent for DLA  column densities.  As we will see later, the
significant difference in the predicted neutral fraction  for sub-DLA
column densities for the two self-shielding models  is important for
our estimates of the metallicity [O/H] of the observed \oi absorbers. 

In the second to  fourth panels of Figure \ref{selfshielding}, we show
the effect  of  the two self-shielding models on the distribution of
neutral hydrogen. The threshold self-shielding model is only shown for
our fiducial value $\log (\Gamma / \textnormal{s}^{-1}) = -12.8$. The \citet{rahmati2013}
self-shielding model was applied to the simulation for two different
assumed values of the photo-ionization  rate,  $\log (\Gamma /
\textnormal{s}^{-1}) = -12.8$ and $\log (\Gamma / \textnormal{s}^{-1})
= -14.0$, respectively. 

As $\Gamma$ decreases, the density threshold for self-shielding
decreases  and the fully neutral self-shielded regions fill an
increasing fraction of the volume  in the simulations. The
self-shielded regions  ``move out'' from the  outer part of galaxy
haloes  into the filaments \citep{miralda2000}. 
As discussed by BH13,
the increase  of the volume filling factor of self-shielded regions
directly translates into an increase of the expected incidence rate of
\hi absorbers optically thick to ionizing radiation. As we will see
later this  can also  explain  nicely why \citet{becker2011} observed
an increasing incidence of \mbox{O\,{\sc i}} systems with increasing
redshift. Note that, over this redshift interval, the incidence of self-shielded regions may be impacted far more strongly by changes in the photo-ionization rate than by the evolution of the density field.

\section{Comparing simulated  and observed O\,{\sevensize\bf I} absorbers} 

\subsection{Synthetic O\,{\sevensize\bf I} absorption spectra}

Using  sightlines extracted from the simulation, synthetic
\mbox{O\,{\sc i}} spectra were generated by first calculating the
optical depth and hence the normalised flux. The optical depth at each
pixel was calculated using the temperature of the gas, its peculiar
velocity, the number density of  \mbox{H\,{\sc i}} and assuming a
relationship $n_{\textnormal{\scriptsize{OI}}} = Z_{\rm O} \, f_{\textnormal{\scriptsize{OI}}/\textnormal{\scriptsize{HI}}} \,
n_{\textnormal{\scriptsize{HI}}}$. Here, $Z_{\rm O}$ is the
metallicity of the gas which is applied to the simulation in
post-processing and $f_{\textnormal{\scriptsize{OI}}/\textnormal{\scriptsize{HI}}}$ is the ratio of the neutral fractions 
of \oi and \hi which we estimated using the photo-ionization code
CLOUDY \citep{ferland1998} as described in the appendix.  
The solar abundance of oxygen was taken as $\log
Z_{\rm O, \odot} = -3.13$ \citep{asplund2009}.  The effect of
instrumental broadening was included by convolving the spectra with a
Gaussian with a FWHM of 6.7 km s$^{-1}$, equal to the instrumental
profile of HIRES (which accounts for  the majority of the
\mbox{O\,{\sc i}} detections in the \citet{becker2011} sample.)

\begin{figure*}
\includegraphics[width=2.0\columnwidth]{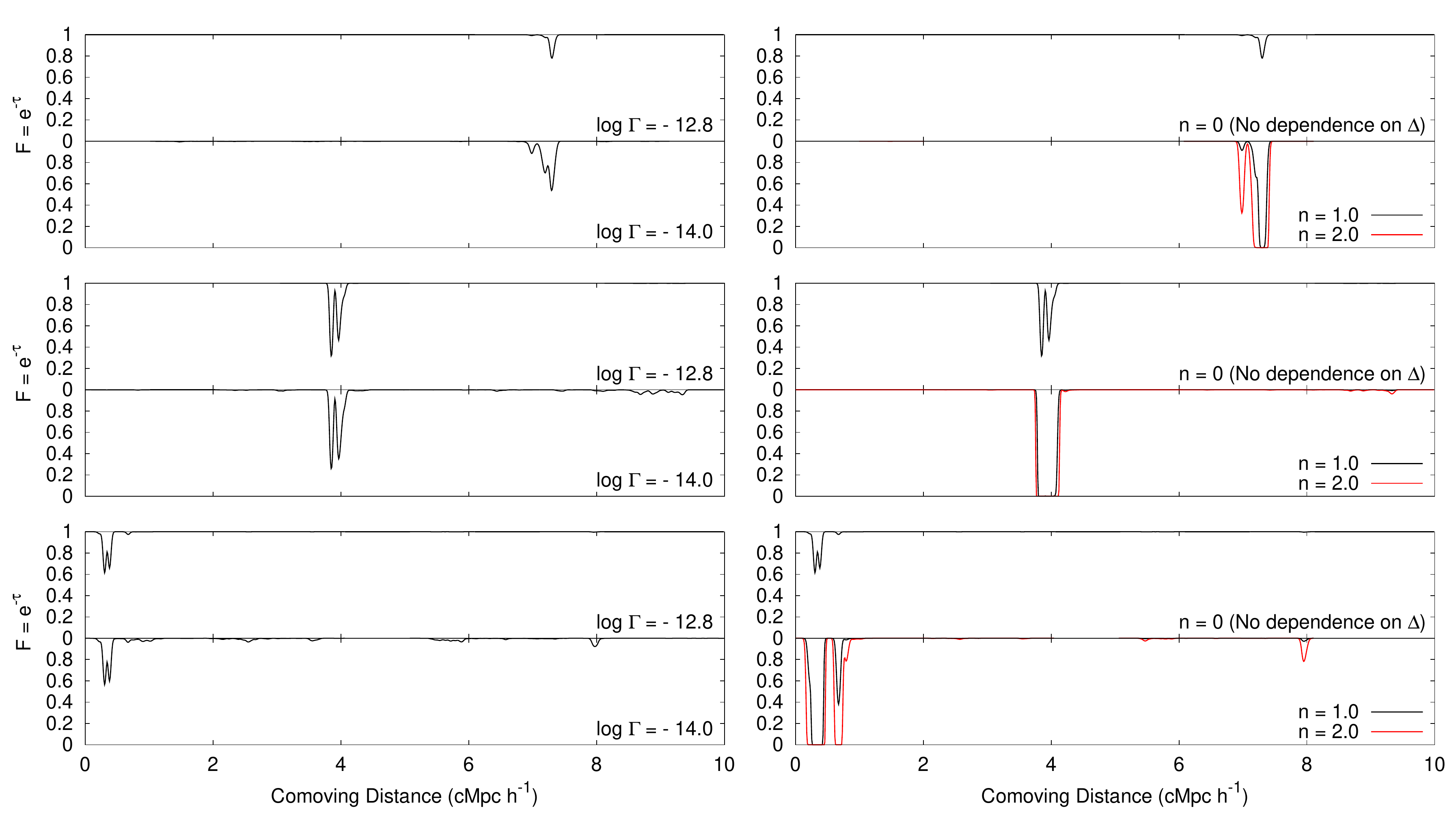}
\caption{Left: Comparison of simulated \oi spectra for two different
  background photo-ionization rates for three different sightlines
  through the simulation at $z=6.0$. The \citet{rahmati2013}
  self-shielding model was used. The top spectrum has $\log (\Gamma /
  \textnormal{s}^{-1}) = -12.8$ and the bottom spectrum has $\log
  (\Gamma / \textnormal{s}^{-1}) = -14.0$. The metallicity is $Z =
  10^{-2} \, Z_{\odot}$. Right: Comparison of simulated \oi spectra
  for different values of the power law index $n$ of the metallicity
  model. The \citet{rahmati2013} self-shielding model was used with
  background photo-ionization rate $\log (\Gamma /\textnormal{s}^{-1})
  = -12.8$.}
  \label{oi_spectra}
\end{figure*}

The left panel of Figure \ref{oi_spectra} shows the effect of
changing $\Gamma$ on the \mbox{O\,{\sc i}} spectra for three different
sightlines. The spectra were generated assuming   $\log (\Gamma /
\textnormal{s}^{-1}) = (-12.8, -14.0)$. As expected, there are more
absorption lines in the  spectra simulated assuming a smaller
$\Gamma$,  due  to the increasing volume filling factor of fully
neutral regions seen in  Figure \ref{selfshielding}.

\subsection{Modelling the metallicity as a function of overdensity}

To model the \oi absorption, we also need to make assumption for the
metallicity [O/H] of the absorbing gas. To do this from first
principles is difficult as this requires correctly modelling where and
how efficiently stars form, as well as the metal yields and the
transport and mixing of metals out of the galaxies into   the
circumgalactic and intergalactic media (see \citet{finlator2013} for a
recent attempt with regard to \oi absorption). The density range
probed by  the absorbers is rather limited and we will make here the
simple assumption  of a power-law  dependence of the  metallicity, $Z
\propto \Delta^n$, without any scatter.  This is certainly a rough
approximation but, as we will see  later, it allows us to reproduce
the \oi absorbers remarkably well.  We normalize this  power-law model
at an overdensity in  the middle of the range probed by the simulated
\oi absorbers.

\begin{figure*}
\includegraphics[width=2.2\columnwidth]{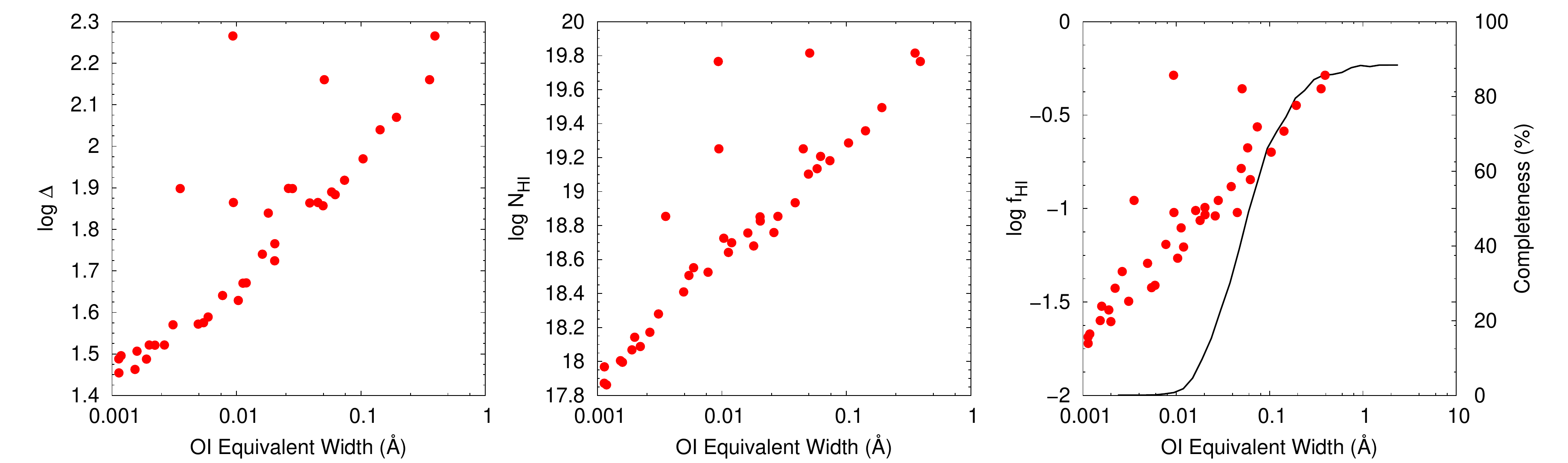}
\caption{Left: The maximum overdensity probed by  simulated \oi
  absorbers as a function of their equivalent width for a model  that
  reproduces the incidence rate observed by \citet{becker2011} at
  $z=6.0$.  The
  \citet{rahmati2013} model for self-shielding was used in each of the
  three panels. Middle: The \mbox{H\,{\sc i}} column density of
  simulated \oi absorbers as a function of their  equivalent
  width. Right: The circles are the neutral hydrogen fraction  of
  simulated \oi absorbers  as a function of their  equivalent
  width. An estimate of the completeness of the observed sample  of
  \oi absorbers by \citet{becker2011} as a function of their
  equivalent width is shown as the black curve  with values on the
  right hand side axis of the panel.   The completeness estimate does
  not reach 100 per cent as some regions of the quasar absorption
  spectra are contaminated due to atmospheric absorption or lines from
  other atomic transitions.}
  \label{od_cd_comp}
\end{figure*}

The left  panel of Figure \ref{od_cd_comp} shows the range of
overdensities   probed by the simulated \oi absorbers for a model that
fits the incidence rate observed by \citet{becker2011}. 
The overdensities probed  range from $1.4 < \log \Delta < 2.3$ at
$z=6.0$. An  overdensity $\Delta = 80$ falls approximately in the
middle of this range and was used as pivot point for the power-law
model of the metallicity. The metallicity was thus parameterized as, 
\begin{equation}
Z = Z_{80} \, \left(\frac{\Delta}{80}\right)^{n},
\end{equation}
where $Z_{80}$ is the metallicity at $\Delta = 80$ and $n$ is the
power-law index. The right panel of Figure \ref{oi_spectra} shows the
effect on the \oi spectra of varying this power-law index  for $n =
(0,1,2)$.

The \mbox{H\,{\sc i}} column densities probed by the simulated \oi
absorbers  are shown in the middle panel of  Figure \ref{od_cd_comp}.
They range from $17.8 < \log N_{\textnormal{\scriptsize{HI}}} <
19.9$. As we will discuss in more detail in the appendix 
\oi is  a good tracer of \mbox{H\,{\sc i}}  if the gas is
``well-shielded''. The right panel of Figure \ref{od_cd_comp} shows the neutral hydrogen
fraction of the \oi absorbers, which ranges from 10 percent to fully
neutral for absorbers with EW $> 0.01 \, \rmn{\AA}$.

\subsection{Comparison to observations}

To compare the simulated  \mbox{O\,{\sc i}} absorption  to the
\citet{becker2011} sample, the equivalent widths (EWs) of the lines
were measured. A threshold flux was calculated and regions where the
calculated flux was lower than this threshold were identified as
absorption features. The equivalent width per pixel was then
calculated and summed over the number of pixels spanned by the
absorption feature to calculate the total equivalent width.

 \citet{becker2011} have  quantified the effect of noise on the
 detection probability of their \oi absorbers, so the equivalent
 widths were measured for  synthetic spectra with no added noise.  An
 estimate of the completeness of the sample as a function of
 equivalent width as determined by \citet{becker2011} for their
 observed sample  is shown as the solid curve in the right panel of
 Figure \ref{od_cd_comp} (with values on the right hand side axis).

\begin{figure*}
\includegraphics[width=2.2\columnwidth]{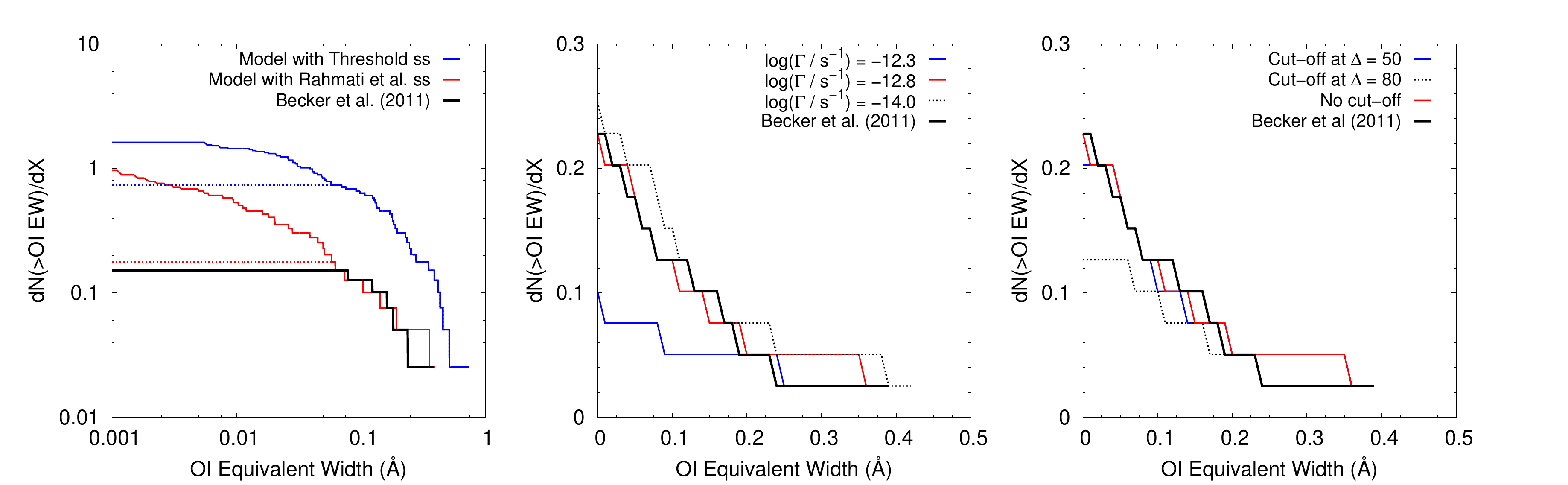}
\caption{The cumulative incidence rate of simulated  \oi absorbers
  compared to the observed distribution  obtained by
  \citet{becker2011}. The left panel is for  the metallicity model
  that fits the data with the \citet{rahmati2013} self shielding model
  ($Z =   10^{-2.65} \, Z_{\odot} \, (\Delta / 80)^{1.3}$).  The
  dashed/solid  curves are  with/without the Becker et
  al. completeness correction.  The middle panel shows the cumulative
  incidence rate of simulated \oi absorbers   assuming different
  background photo-ionization   rates $\Gamma$  and the
  \citet{rahmati2013} self-shielding model. The completeness
  correction shown in the right panel of Fig. \ref{od_cd_comp} has
  been applied to the curves. The metallicity model used is the same
  as in the left panel. The right panel  shows the effect of
  introducing  different cut-off overdensities, below which the
  metallicity is set to zero. The red curve assumes a power law model
  for the metallicity that extends to  arbitrarily low density. The
  blue curve assumes that  regions with $\Delta < 50$  contain no
  metals. The black dashed curve assumes that there are no metals
  below  a cut-off overdensity of   $\Delta = 80$. The thick black
  curve shows  the observed cumulative incidence rate of \oi absorbers
  obtained  by \citet{becker2011}. The metallicity model used is the
  same as in the left panel.}
  \label{cp_comp_gamma}
\end{figure*}

For a quantitative comparison with the simulated \oi absorption to the
\citet{becker2011} sample,  we have compiled cumulative incidence
rates as shown  in Figure \ref{cp_comp_gamma}.  The parameters of the
assumed metallicity density relation  were  chosen to give a good
match to the data for our fiducial  photo-ionization rate $\log (\Gamma /
\textnormal{s}^{-1})= -12.8$, a value consistent with the observations
at $z=6$ as shown in the right panel of   
Figure \ref{survey_hfrac_gamma} \citep{calverley2011, wyithe2011}. The left
panel of Figure \ref{cp_comp_gamma} shows the cumulative distribution
with and without applying the completeness correction  to the sample
of simulated \oi spectra. As the left panel shows,  a
metallicity-density relation with  $Z_{80} = 10^{-2.65} \, Z_{\odot}$
gives a good match to the data for the \citet{rahmati2013}
self-shielding model. 

We have also tested the effect of using the threshold density for
self-shielding instead of the  \citet{rahmati2013} prescription.  With
the threshold self-shielding model, self-shielded regions 
have a larger covering factor. There are therefore  many more \oi 
absorbers of a given  EW
and the EW distribution extends to larger values for the same
metallicity.  It is possible  to fit the  observed EW distribution of
the observed \oi absorbers   with either of the two self-shielding
prescriptions we have implemented,  but, as we discuss in more detail
in section 4.1, the inferred metallicity differs significantly.  As
the \citet{rahmati2013} prescription includes radiative transfer
effects, we will  consider our simulations with this prescription to
be our fiducial self-shielding model.

In the middle panel of Figure \ref{cp_comp_gamma}, we show how
changing $\Gamma$  affects the cumulative incidence rate. The  number
of lines seen at small equivalent widths increases  strongly as
$\Gamma$ decreases, but the number of lines seen at larger equivalent
widths is insensitive to changes in $\Gamma$. We have tested a wide
range  of  $\Gamma$. The incidence rate of weak \oi absorbers
decreases rapidly   with increasing   photo-ionization rate, compared
to our fiducial value of $\log (\Gamma / \textnormal{s}^{-1}) =
-12.8$. When we decrease  the photo-ionization rate, the incidence
rate quickly saturates  once the \oi absorbers have become fully
neutral. We show here two  extreme cases with  $\log (\Gamma /
\textnormal{s}^{-1}) = (-12.3, -14.0)$, which represent a highly
ionized and significantly neutral IGM, respectively.

The right panel of Figure \ref{cp_comp_gamma} shows the effect of
applying a cut-off to the metallicity at different
overdensities. Below this cut-off, it is assumed that no metals are
present. Above this cut-off, the metallicity follows the power-law
model as before.  The simulations of 
\citet{finlator2013} which attempt to model the  metal transport into
low density regions predict such a cut-off.  The details of this
cut-off  will, however,  depend sensitively  on  the details of the
galactic wind implementations in numerical simulations. 
A cut-off of $\Delta = 80$ (0.003\% of simulation
volume) produced too few simulated \oi absorbers compared to the
\citet{becker2011} survey.  A cut-off of $\Delta = 50$ (0.01\% of
simulation volume) made only a small difference compared to not using
a cut-off at all. This suggests that the metals have travelled out to
regions with overdensities  as low as $\Delta \sim  50-80$  by $z \sim 6$ and that the 
metallicity at lower densities is not (yet) probed by the current data. 
Note that the effect  of a high  cut-off overdensity is similar to that of
a high $\Gamma$.

We have also compared  the  velocity widths, $\Delta v_{90}$,  of the
simulated \oi absorption to that of the \citet{becker2011} survey.
For this we have  identified the velocity interval covering 90\% of
the optical depth in the \oi absorption systems.  The velocity widths
of our simulated \oi absorbers  range from $18.5 \, \rm km \, \rm
s^{-1}$ for a line with equivalent width $W_{\lambda} = 0.02 \,
\textnormal{\AA}$ to  $45.0 \, \rm km \, \rm s^{-1}$ for a line with
equivalent width $W_{\lambda} = 0.06 \,  \textnormal{\AA}$. These
$\Delta v_{90}$ values are similar to  the velocity widths measured by
\citet{becker2011} for their observed  \mbox{O\,{\sc i}} absorption
lines.

\begin{figure*}
\includegraphics[width=2.2\columnwidth]{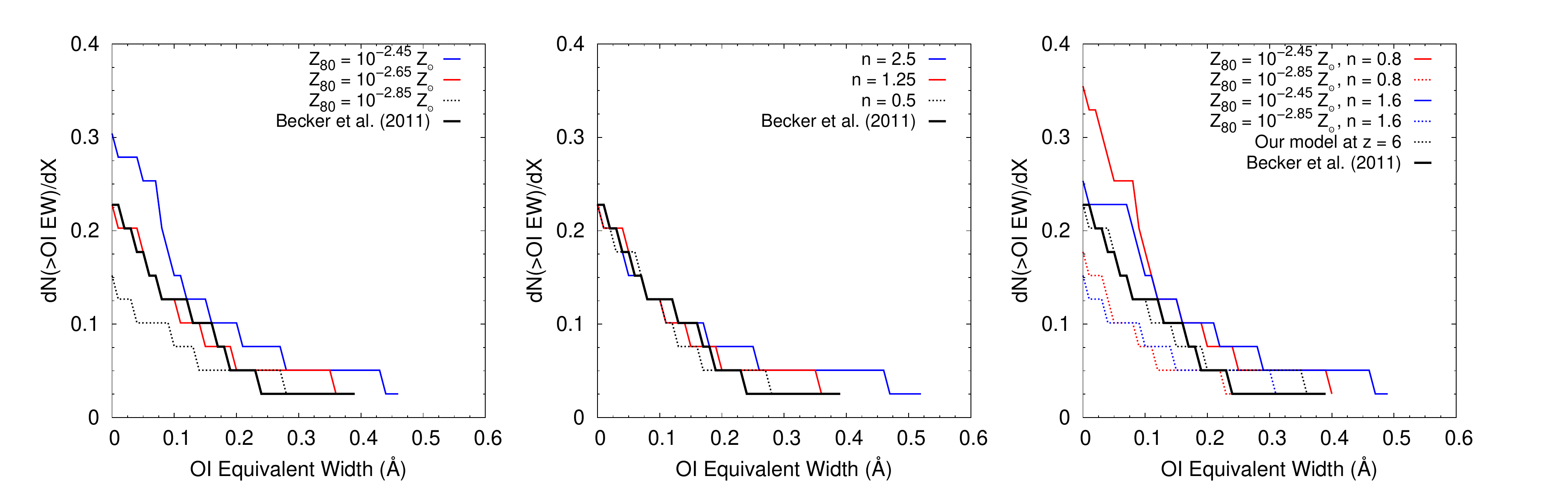}
\caption{The cumulative incidence rate of \oi absorbers  showing the
  effect of changing  slope and normalization of the  metallicity
  model ($Z = Z_{80} \, (\Delta / 80)^{n}$). All curves shown have a
  completeness correction applied.  The assumed photo-ionization rate
  for the simulated \oi absorbers  is $\log (\Gamma /
  \textnormal{s}^{-1}) = -12.8$ and the \citet{rahmati2013}
  self-shielding model was used. The left panel shows the effect of
  varying the normalisation $Z_{80}$. The middle panel shows the
  effect of changing the power law index $n$. The right panel shows
  the four metallicity models that define the blue-shaded region in
  the left panel of Figure \ref{metallicity_lls}.}
\label{cp_parameters}
\end{figure*}

Varying $Z_{80}$ and $n$ in the model of the metallicity, as well as
changing $\Gamma$, both have very  noticeable effects on the number of
absorption lines produced by the simulation. 
As shown in Figure \ref{cp_parameters},  
increasing the normalisation $Z_{80}$ (left
panel) does not change the slope of the cumulative distribution while
varying the power law index $n$ does.  The maximum equivalent width
seen in the simulation  decreases with decreasing $n$. While some degeneracy exists between $Z_{80}$ and $n$, therefore, we have some leverage with which to constrain these parameters by matching both the slope and the amplitude of the observed distribution.

\section{Results}

\subsection{The metallicity and photo-ionization rate  at $z \sim 6$}

\begin{figure*}
\includegraphics[width=2.2\columnwidth]{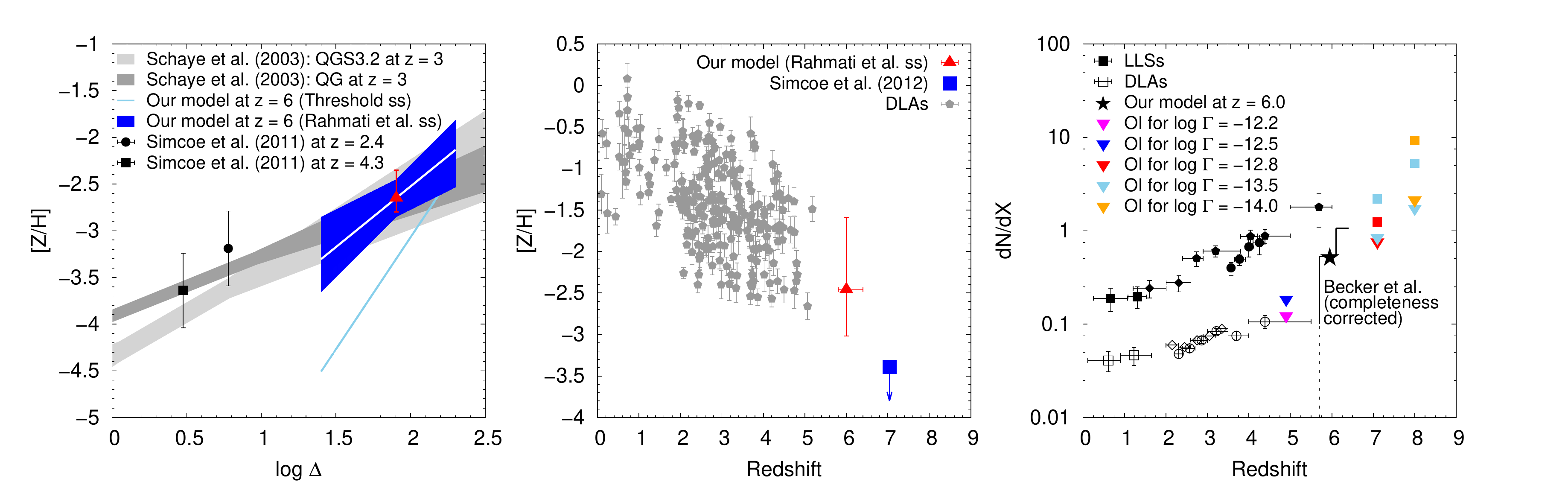}
\caption{Left: The dark  blue shaded region and the light blue line
  show the metallicity  inferred as a function of overdensity  for the
  \citet{rahmati2013}   and the threshold self-shielding model,
  respectively.  The dark blue shaded  region is thereby  defined  by
  the four metallicity models  shown in the right panel of Figure
  \ref{cp_parameters} and the central white line is our fiducial
  metallicity model.  The assumed  photo-ionization rate is  $\log
  (\Gamma / \textnormal{s}^{-1})  = -12.8$ and the value of the
  inferred $Z_{80}$ for our fiducial metallicity model  is denoted  by
  the red triangle. The  red bar shows how the inferred $Z_{80}$
  increases using the \citet{rahmati2013} self-shielding model as the
  photionization  rate is increased in the range $\log (\Gamma /
  \textnormal{s}^{-1})  = (-13.5, -12.5)$. Also shown is the
  metallicity as a function of overdensity at $z=3$, as measured by
  \citet{schaye2003},  in grey for two different models of the UV
  background and measurements of the metallicity of the IGM by
  \citet{simcoe2011}.  Middle: The red triangle shows the median
  $z\sim 6$ metallicity of completeness-corrected simulated \oi absorbers with EW$_{\textnormal{\scriptsize{OI}}} \ge 0.01$ for our  fiducial
  model as measured by comparing total \oi and \hi column density.
  The vertical error bar represents the $1\sigma$ range of
  metallicities of the simulated absorbers, while  the horizontal
  error bar indicates the redshift range of the  \oi absorbers
  observed by \citet{becker2011}. The grey points show a compilation
  of  metallicity measurements of DLAs compiled in
  \citet{rafelski2012}. The upper limit for the metallicity measured
  by \citet{simcoe2012} for a possible proximate  DLA in the
  foreground of the $z=7.085$ QSO ULASJ1120+0641 is also shown. Right:
  Comparison of the predicted evolution of the  incidence rate ${\rm
    d}N/{\rm d}X $ of our simulated \oi absorbers with EW$_{\textnormal{\scriptsize{OI}}} \ge 0.01 \, \rmn{\AA}$ and and that of
  observed  LLSs (black points) and  DLAs (open points). The square points at $z=7$ and $z=8$  show predictions for the 
incidence rate of OI absorbers with a threshold EW$_{\textnormal{\scriptsize{OI}}} = 0.001 \, \rmn{\AA}$. LLSs:
  \citet{songaila2010} and as compiled in \citet{fumagalli2013}. DLAs:
  As compiled in \citet{seyffert2013}.}
\label{metallicity_lls}
\end{figure*}

In the left panel of Figure \ref{metallicity_lls} we show the
metallicity-density relation  of our fiducial model which reproduces
the incidence rate of the observed \oi absorbers and compare it to
that measured   by \citet{schaye2003} at $z = 3$ for two plausible
models of the UV background at this redshift\footnote{ Model QG is the
  \citet{haardt2001} used in our simulation, which includes
  contributions from both quasars and galaxies.  Model  QGS3.2 has a
  flux that is 10 times smaller above 4 ryd than model QG in order to
  mimic the effect of incomplete reionization of helium.},  as well as
to those of \citet{simcoe2011} at $z=2.4$ and $z=4.3$.  Note that
these metallcities have been lowered by 0.12 and 0.09 dex,
respectively, to match the \citet{asplund2009} measurements of solar
abundances assumed here.

Our fiducial metallicity density relation, utilizing the
\citet{rahmati2013} self-shielding model, is given by,  
\begin{equation}
Z=10^{-2.65} \, Z_{\odot} \left(\frac{\Delta}{80}\right)^{1.3},
\end{equation}
with a background photo-ionization rate $\log (\Gamma /
\textnormal{s}^{-1}) = -12.8$.  This relation was determined by
calculating ${\rm d}N(>{\rm EW}_{\rm OI})/{\rm d}X$ and the maximum
\oi EW for a range of $Z_{80}$ and $n$ and finding the best match to
the observations of \citet{becker2011} with  a K-S test.

Rather surprisingly,  our modelling of the \oi absorbers suggests that
there is little, if any,   evolution of the metallicities of the CGM
between $3<z<6$ in the overdensity range probed by the \oi absorbers.
Note, however, that our assumption of  no scatter in the metallicity
density relation is certainly not realistic. We will come back to this
later.  For reference, we also show the inferred metallicities for the
threshold self-shielding model. As already  discussed, the
metallicities are typically a factor of ten lower (with a steeper
density dependence) due to the larger neutral fractions  in this
model. 

When varying the metallicity distribution and photo-ionization rate,
we found a rather  weak  dependence of the \oi incidence rate on the
latter, which was degenerate with adjusting the metal distribution
unless the photo-ionization was so high that  the gas in self-shielded
region became highly ionized and their  incidence rate  too low to
reproduce the \citet{becker2011} data.  That occurred at  $\log \Gamma
\ga  -12.4$.  The red bar in the left panel of Figure
\ref{metallicity_lls} shows how the   inferred $Z_{80}$ changes   as
the photo ionization rate varies in the range  $-13.5< \log (\Gamma /
\textnormal{s}^{-1}) < -12.5$.  As expected, the inferred metallicity
thereby increases with increasing photo-ionization rate (the value
denoted by the red triangle  is for  our fiducial model with  $\log
(\Gamma / \textnormal{s}^{-1}) = -12.8$ suggested by measurements 
of the photo-ionization rate from \lya forest data).  
The dependence of the inferred metallicity on the assumed 
photo-ionization rate is thereby weak (a change of 0.5 dex 
in inferred metallicity for a change of 1.6 dex of the photo-ionization
rate).  We also note again that modelling the self-shielding
carefully is important. With the simple threshold self-shielding model 
the normalization of the inferred metallicity at the characteristic
overdensity $\Delta =80$ would be 0.8 dex lower and the inferred
dependence on overdensity would be significantly steeper than for the 
\citet{rahmati2013}  self-shielding model.

The number of observed  \oi absorbers is still far too small  for a
robust  determination of the differential  \oi EW distribution and its
errors, but as we will discuss in more detail in section  4.5, the
main uncertainties in the inferred metallicity are anyway  due to
still uncertain model assumptions as e.g. demonstrated  in Figure
\ref{metallicity_lls} by the  sensitivity to the details of the
self-shielding model and the assumed photo-ionization rate.  We have
thus not attempted to calculate formal confidence intervals for
$Z_{80}$ and $n$.  The  blue shaded region in the left panel of Figure
\ref{metallicity_lls}  shows instead a   range of metallicity density
relations at fixed photo-ionization rate   for which we show the
corresponding cumulative \oi  incidence rate  in the right panel of
Figure \ref{cp_parameters}.

In the middle panel of Figure \ref{metallicity_lls},  we compare the
metallicity that would be measured  along lines of sight through our
simulation  that contain \oi absorbers  by comparing the total  \oi
and \hi column densities with the measured metallicity of DLAs  at a
range of redshifts\footnote{\citet{wolfe1994}, \citet{meyer1995},
  \cite{lu1996}, \citet{prochaska1996}, \citet{prochaska1997},
  \citet{boisse1998}, \citet{lu1998}, \citet{lopez1999},
  \citet{pettini1999}, \citet{prochaska1999b}, \citet{churchill2000},
  \citet{molaro2000}, \citet{petitjean2000}, \citet{pettini2000},
  \citet{prochaska2000}, \citet{rao2000}, \citet{srianand2000},
  \citet{dessauges2001}, \citet{ellison2001}, \citet{molaro2001},
  \citet{prochaska2001a}, \citet{prochaska2001b}, \citet{ledoux2002a},
  \citet{ledoux2002b}, \citet{levshakov2002}, \citet{lopez2002},
  \citet{petitjean2002}, \citet{prochaska2002}, \citet{songaila2002},
  \citet{centurion2003}, \citet{ledoux2003}, \citet{lopez2003},
  \citet{prochaska2003a}, \citet{prochaska2003b},
  \citet{dessauges2004}, \citet{khare2004}, \citet{turnshek2004},
  \citet{kulkarni2005}, \citet{akerman2005}, \citet{rao2005},
  \citet{dessauges2006}, \citet{ledoux2006}, \citet{meiring2006},
  \citet{peroux2006}, \citet{rao2006}, \citet{dessauges2007},
  \citet{ellison2007}, \citet{meiring2007}, \citet{prochaska2007},
  \citet{nestor2008}, \citet{noterdaeme2008}, \citet{peroux2008},
  \citet{wolfe2008}, \citet{jorgenson2010}, \citet{meiring2011},
  \citet{vladilo2011}, \citet{rafelski2012}.}. Our inferred
metallicity for the \oi absorbers  is at the lower end of the range of
measured DLAs  at somewhat lower  redshift. This should not be
surprising  given the sub-DLA column densities we have inferred for
the \oi absorbers which should probe the CGM at somewhat  lower
overdensities and larger distances from the host galaxy  than the
DLAs. We also show the  upper limit for the metallicity  obtained by
\citet{simcoe2012} for the case that the  damping wing redwards of
\lya  in the $z=7.085$ QSO ULASJ1120+0641 is interpreted as due to a
proximate foreground DLA   at  $z=7.041$. This  upper limit is
significantly  below our estimate  for the metallicity of the \oi
absorbers. We have also calculated  predicted metallicities for absorbers 
with $\log N_{\textnormal{\scriptsize{HI}}} \ge 20.45$ (the column density range inferred by \citet{simcoe2012}) for our fiducial metallicity model and 
photo-ionization rates $-14.0 \le  \log \Gamma \le 12.8$ and found them
to be a factor three or more higher than the upper limit on the
metallicity inferred by  \citet{simcoe2012}. This renders  the interpretation 
as a proximate  foreground DLA rather unlikely (see also BH13 and the 
corresponding discussion in \citet{finlator2013}  who come to similar 
conclusions), but we should note again here that we made no attempt 
to model the likely scatter in metallicity of the CGM in our simulations.

\subsection{The relation to lower redshift DLAs and LLSs}

In the right panel of Figure \ref{metallicity_lls}, we compare the
redshift evolution of the incidence rate of our modelled \oi absorbers
for a range  of plausible  assumptions of the photo-ionization rate
to the evolution of the incidence of LLSs and DLAs at lower
redshift\footnote{LLSs: \citet{songaila2010}, \citet{omeara2013},
  \citet{ribaudo2011}. DLAs: \citet{rao2006}, \citet{prochaska2009},
  \citet{noterdaeme2012}.}. As already discussed by
\citet{becker2011}, the incidence rate of their observed  \oi
absorbers  at $z=6$ is similar to that of LLSs at $z=4$. 
The incidence rate of our simulated \oi absorbers at $z=5$  matches
very well  with that of observed  DLAs and LLSs at lower redshift.
At $z> 5$ the photo-ionization rate appears to drop 
(right panel of  Figure \ref{survey_hfrac_gamma}; \citealt{wyithe2011};
\citealt{calverley2011}) which   can be mainly attributed to a rapidly
decreasing mean free path with increasing redshift as the tail-end of
reionization is approached \citep{mcquinn2011}.  
If we assume such a drop of the photo-ionization rate
our simulations  reproduce the rapid  evolution of the incidence rate of
the \oi absorbers in the \citet{becker2011} data very well. 
As we will discuss in more detail in the next section, the  rapid
evolution of the incidence rate of the simulated \oi absorbers
is expected to continue at $z>6$. Inspection of  Figure  \ref{selfshielding}  
further shows that  the incidence rate depends more strongly on 
the photo-ionization rate and only to a lesser extent  on  the
increasing density with redshift. This explains {\it e.g.}  the rather small
difference in the incidence rates of our modelled  \oi absorbers at 
$z=6$ and $z=7$ for our fiducial photo-ionization rate ($\log \Gamma = -12.8$)
shown by the star and the red triangle in the right panel of
\ref{metallicity_lls}, respectively.

\subsection{The spatial distribution of \oi absorbers}

\begin{figure*}
\includegraphics[width=2.0\columnwidth]{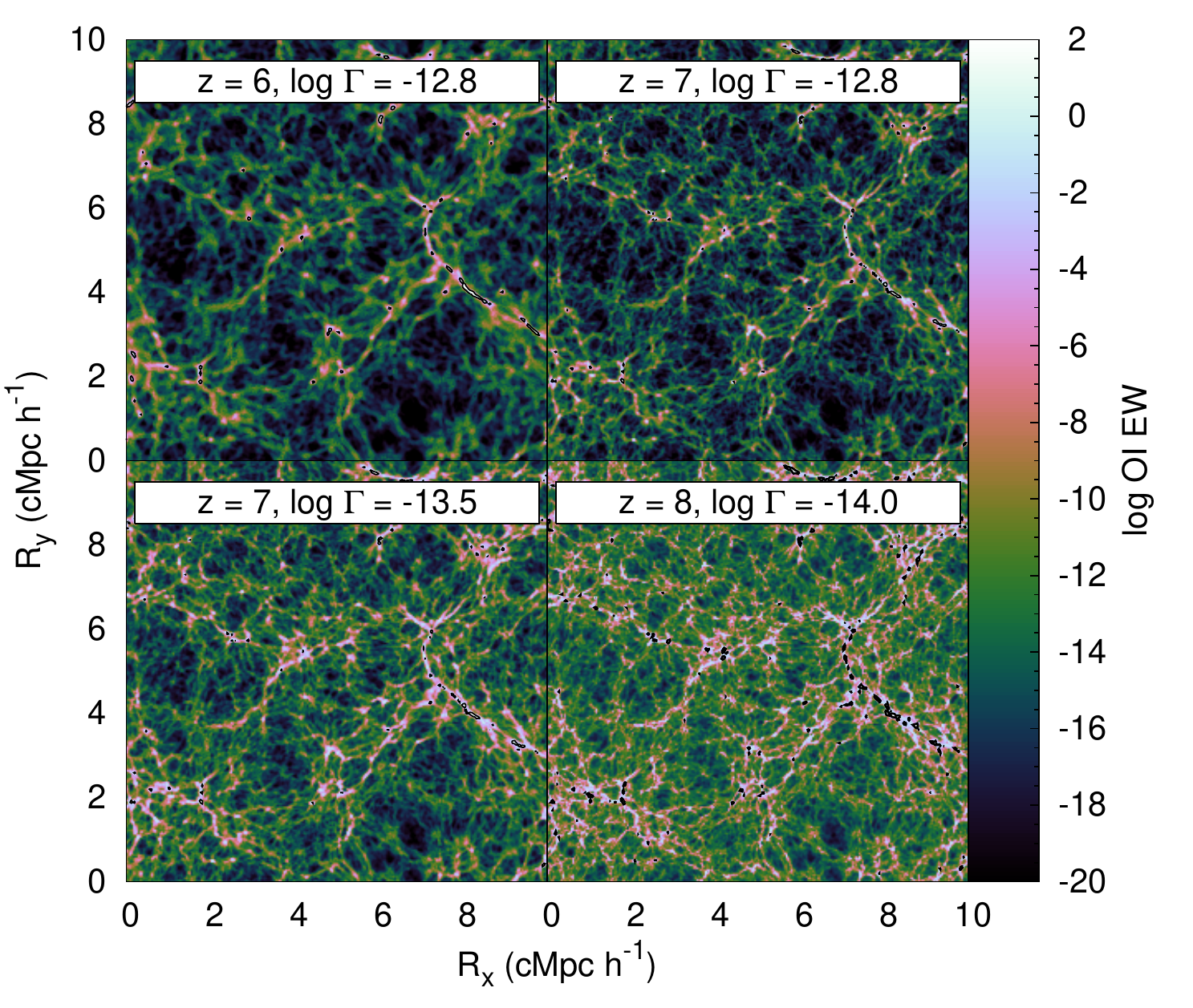}
\caption{The location of \oi absorbers in the simulation. The \oi
  equivalent width was estimated from the  \hi column density (see
  Figure \ref{od_cd_comp}). The \hi column density was calculated by
  integrating $n_{\rm HI}$ over the thickness ($39h^{-1}
  \,\rmn{ckpc}$) of the  slice shown in the plot.  The distribution of
  \oi absorbers is shown for $\log \Gamma = -12.8$ at $z = (6.0,
  7.1)$, for $\log \Gamma = -13.5$ for $z=7.1$ and for  $\log \Gamma =
  -14.0$ at $z=8.0$. The \citet{rahmati2013} model for self-shielding
  was used in each case. The black contours enclose  absorbers with 
   ${\rm EW}_{\scriptsize{\rmn{OI}}}\ge 0.01 \, \rmn{\AA}$.}
  \label{absorber_dist}
\end{figure*}

Figure \ref{absorber_dist} shows the spatial distribution of  the \oi
EW for our models for a range of redshifts and  photo-ionization
rates.  For this we have calculated \hi  column densities  by
integrating the neutral hydrogen  density as shown in Figure
\ref{selfshielding}    with the \citet{rahmati2013} self-shielding
model  over the thickness of the slice shown and used the  correlation
between \hi column density and \oi EW  in the middle panel of Figure
\ref{od_cd_comp} to translate the \hi column densities  into   \oi
equivalent widths. The relation used was
\begin{equation}
\log \rmn{EW_{OI}} = 2.51 \log N_{\scriptsize{\rmn{HI}}} - 47.86.
\end{equation}
The black contours shows the location of \oi absorbers with 
${\rm EW}_{\scriptsize{\rmn{OI}}}\ge 0.01 \, \rmn{\AA}$.   Figure \ref{absorber_dist} suggests that the observed
absorbers at  $z \sim 6$ indeed probe the (outer parts) of the haloes
of (faint) high-redshift  galaxies as suggested by \citet{becker2011}.
If the metal distribution extends to lower densities than currently 
probed  the \oi absorbers are expected to start to probe
more and more the filamentary structures  connecting these galaxies as
the meta-galactic photo-ionization rate decreases and the  Universe
becomes increasingly more neutral with increasing redshift.

\subsection{Predictions for the incidence rate of \oi absorbers at $z = 7$ and beyond}

We have also used our simulations to predict the evolution of the
number of absorption systems at $z=7$ and $z=8$. For this we assumed
the inferred metallicity-(over)density relation  at $z=6$.  This seems
to be a reasonable  assumption given the apparent absence of a
metallicity  evolution at the relevant  densities between $z=3$ and
$z=6$.

\begin{figure*}
\includegraphics[width=2.2\columnwidth]{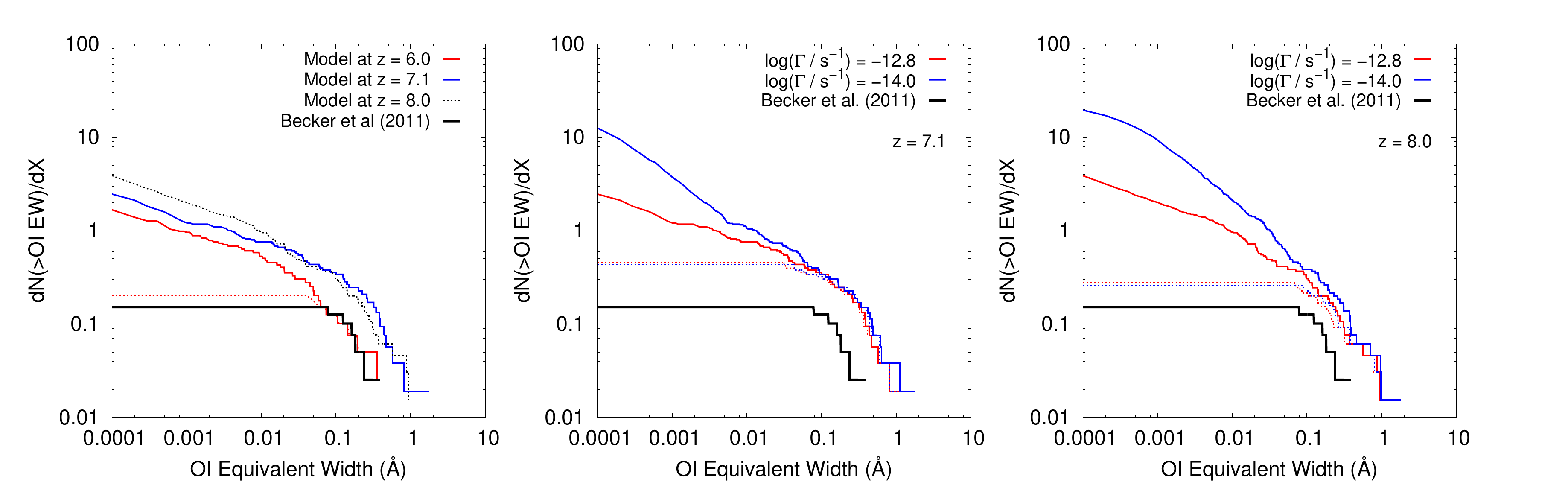}
\caption{Left: The solid red and blue lines and the black dashed line show the predicted cumulative incidence rate of \oi
  absorbers without completeness correction for  the metallicity
  distribution  that fits the observed distribution   well at $z=6.0$
  with the \citet{rahmati2013} self-shielding model and $\log (\Gamma / \textnormal{s}^{-1}) = -12.8$.  The red dashed
  curve is for the simulated \oi absorbers at $z=6.0$  with the  completeness
  correction applied. The thick black curve is the \citet{becker2011}
  data. Middle: The thin solid lines show the predicted cumulative incidence rate of \oi
  absorbers   without completeness correction  for different values of
  the background photo-ionization rate at $z=7.1$. The dashed curves
  show the effect of applying a cut-off in the metallicity below
  $\Delta = 50$. Right:  The thin solid lines show the predicted cumulative incidence rate of
  \oi absorbers  without completeness correction for different values
  of the background photo-ionization rate at $z=8.0$. The dashed
  curves show the effect of applying a cut-off in the metallicity
  model below $\Delta = 50$.}
  \label{cp_redshift}
\end{figure*}

The left panel of Figure \ref{cp_redshift} shows a cumulative plot for
the incidence rate ${\rm d}N(>{\rm EW}_{\rm OI})/{\rm d}X$ for our
modelling  at $z=(6.0,7.1,8.0)$ for a fixed photo-ionization rate,
$\log (\Gamma / \textnormal{s}^{-1}) = -12.8$.  The results for $z=6$
are shown in red with (dotted curve) and without (solid curve)
completeness correction,  while the results at $z=7.1$ and $z=8.0$ are
shown only without the \citet{becker2011} completeness correction  as
the blue solid and black dotted curve, respectively. The significant,
but moderate evolution is here only due to the increasing (column)
densities with increasing redshift.

The two right panels of Figure \ref{cp_redshift} show the cumulative
incidence rate  for two  background photo-ionization rates at $z=7.1$
and $z=8.0$, respectively. The solid lines represent metallicity
models with no cut-off and the dashed lines represent models with a
cut-off for the presence of metals  at $\Delta = 50$. If the presence
of metals should indeed be limited to overdensities $\Delta \ge 50$, the total incidence rate saturates at  ${\rm
  d}N(>{\rm EW}_{\rm OI})/{\rm d}X \sim 0.5$ and there is very little, if any, sensitivity to a decrease in the photo-ionization rate. The EW of
the weakest \oi absorbers is  $0.01 \,\rmn{\AA}$, not much larger than
the weakest absorbers detected by \citet{becker2011} at $z\sim 6$.
This is because, in this case, increasing redshift and/or decreasing
photo-ionization  rate increase only moderately  the covering factor
of self-shielded regions  that contain
metals.  If, instead, the presence of metals follows our assumed
power-law  metallicity-density relation all the way to low densities,
there will be a larger number of additional  weak \oi absorbers and
the total incidence rate increases by about a factor of ten  to ${\rm
  d}N(>{\rm EW}_{\rm OI})/{\rm d}X \sim 10$ for \oi absorbers with EW
$>0.001 \, \rmn{\AA}$.  In this case the weakest absorbers remain
sensitive to the value of the decreasing photo-ionization rate.  Note,
however, that pushing to such small \oi EW will almost certainly
require a 30m  telescope. Note  further that  by assuming that
  the metallicity does not evolve with redshift  our  predictions are
  optimistic and smaller numbers of \oi absorbers  would  be expected
  if the metallicity  of the CGM/IGM were  to decrease over the
  redshift range considered.

\subsection{Caveats} 

In this work, we have made a number of simplifying assumptions which
may have affected our results. Most critical is probably our
approximate treatment of radiative transfer effects, especially for
our predictions towards higher redshift where we move deeper into the
epoch of reionization.

\begin{itemize}

\item
Our treatment of self-shielding is based on the full radiative
transfer simulations of \citet{rahmati2013} and should be reasonably
accurate as long as internal sources within the \oi  absorbers do not
dominate the ionizing flux. \citet{rahmati2013b} have  investigated
the effect of local stellar ionizing  radiation and found that it can
have a significant effect for LLSs and sub-DLAs (see also
\citealt{kohler2007}). This may mean that we have somewhat
underestimated the metallicity necessary to produce the observed \oi
absorbers. 

\item
Our assumption of a fixed metallicity-density relation is also
certainly not correct. From observations of DLAs, we know that the
metallicity scatter in absorbers with larger column density is
significant. We should expect this to be the case also for the \oi
absorbers, which are essentially sub-DLAs. In a self-consistent picture
for the ionizing sources and the production and transport of metals 
one may even expect  an anti-correlation between metallicity 
and the amout of neutral gas present. 

\item Our simulations also neglect the effect of outflows on the gas
distribution. There may thus be overall less gas at the impact
parameters producing  \oi absorbers than our simulations predict and
this may lead to a systematic underestimate of metallicites. Note that 
the simulations by \citet{rahmati2013} did include the effect of
galactic winds.

\item
Somewhat  problematic is also the rather small box size  of our
simulations, which is dictated by the need  to resolve the small scale
filamentary structures which dominate the  absorption signatures in
QSO spectra at high redshift (see \citet{bolton2009} for a detailed
discussion). There should thus also be more scatter in the correlation
between overdensity and hydrogen column density than our simulations
suggest.  

\item
For our predictions towards higher redshift, the neglect of
the expected substantial fluctuations  of the photo-ionization rate is
another issue. The expected number  of \oi absorbers should vary
greatly on scales of the mean free path of ionizing photons, which
becomes  comparable or smaller than the box size of our simulations
at $z\ga 6$.  
\end{itemize}

\subsection{Comparison to other work}

Detailed simulations of \oi absorption at the tail-end of reionization
are still in their infancy. Most notably, \citet{finlator2013} have
recently presented full cosmological radiative transfer simulations
which also follow    the metal enrichment history of the
circumgalactic medium.    The relevant processes (reionization, metal
transport by galactic winds) are difficult to simulate from first
principles and it is perhaps not  surprising that \citet{finlator2013}
were somewhat struggling to reproduce   the observed \oi absorbers by
\citet{becker2011}. As far as we can tell  from comparing their work
to ours, the main difference is  -- as the authors point out
themselves -- that in their simulations  the build-up of the
meta-galactic UV background had progressed  already too far by $z\sim
6$. This then leaves too little neutral   gas in self-shielded
regions.  In contrast to our findings,  \citet{finlator2013} also
predict  a decrease of the incidence rate  of \oi absorbers with
increasing redshift. The difference  here can be traced to the fact
that, in their simulations, the filamentary structure  connecting the
DM haloes hosting their galaxies is not metal enriched and/or
substantially neutral. Finally,  we should note that the metallicity
in their simulation at $z\sim 6$ is about a factor of ten   higher
than we infer at the relevant overdensities at the same redshift perhaps suggesting that the implementation of galactic winds in
  their  simulations does not  transport metals to sufficiently large
  distances.

\section{Summary and conclusions}

We have used cosmological hydrodynamical simulations to model the
neutral hydrogen and the associated  OI absorption in  regions of the
circumgalactic medium at high redshift. A simple power-law
metallicity-density relation was  assumed. The self-shielding to
ionizing radiation was modelled by  post-processing  the simulations
using two different schemes. Our simulations reproduce the incidence
rate of  observed \oi absorbers at $z=6$ and their cumulative
incidence rate   with a metallicity [O/H] $\sim -2.7$ at a typical
overdensity $\Delta=80$,  with the presence of oxygen extending to an
overdensity at least as low as $\Delta \sim 50-80$, a moderate density
dependence with power-law index $n=1.3$,    and a photo-ionization
rate in agreement with measurements from \lya forest data, $\log
(\Gamma / \textnormal{s}^{-1}) = -12.8$. The metallicity we infer
here is very similar  to that inferred by \citet{schaye2003}  at $z
\sim 3$ from \civ absorption by (highly ionized)  gas  at a similar
overdensity. There appears, therefore, to be remarkably little
evolution of the typical metallicity  of the CGM between $3<z<6$ at
these overdensities.    The efficient metal enrichment of the CGM
appears to have  started early in the history of the Universe and in
low-mass galaxies.   Our inferred   metallicity is also in good
agreement with the lower end of those of DLAs at slightly lower
redshift. This gives further support to a picture where the observed
\oi absorption arises at  somewhat larger impact parameter and lower
overdensity  with somewhat  lower column density than typical DLAs. 

Our simulations further reproduce the observed rapid redshift
evolution of observed \oi absorbers at $z>5$ for reasonable
assumptions for the evolution of the meta-galactic photo-ionization
rate. The rapid evolution is mainly due to the  self-shielding
threshold moving out first from the inner part into the  outer part of
galactic haloes and then into the filamentary structures of the cosmic
web. This is  due to the decreasing photo-ionization rate as we
progress deeper into the epoch of reionization with increasing
redshift. The observed evolution of the incidence rate of \oi
absorbers  thereby matches well onto that of LLSs and DLAs at lower
redshift.

Finally, we have made predictions for the expected number  of \oi
absorbers at redshifts larger than currently observed and predict that
the rapid evolution will continue with increasing redshift  as the \oi
absorbers probe the increasingly neutral cosmic web.   Pushing the
detection of \oi absorbers to higher redshift ($z>6$) and  lower EW
should therefore provide a rich harvest and should allow  unique
insight into the enrichment history of the circumgalactic medium and
the details of how reionization proceeds.

\section*{Acknowledgments}

The hydrodynamical simulations used in this work were performed using
the Darwin Supercomputer of the University of Cambridge High
Performance Computing Service (http://www.hpc.cam.ac.uk/), provided by
Dell Inc. using Strategic Research Infrastructure Funding from the
Higher Education Funding Council for England.  We thank Volker
Springel for making GADGET-3 available. The contour plots presented in
this work use the cube helix colour scheme introduced by
\cite{green2011}. JSB acknowledges the support of a Royal Society
University Research Fellowship. GDB acknowledges support from the
Kavli Foundation and the support of a  STFC Rutherford fellowship. LCK
and MGH acknowledge support from the FP7 ERC Advanced Grant
Emergence-320596. LCK also acknowledges the support of an Issac Newton
Studentship, the Cambridge Trust and STFC.  This work was further
supported in part by the National Science Foundation  under Grant
No. PHYS-1066293 and the hospitality of the Aspen  Center for
Physics. We thank Len Cowie and Max Pettini for
helpful discussions and suggestions. We also thank Bob Carswell for his advice on the CLOUDY modelling and his helpful comments on the manuscript.

\bibliographystyle{mn2e} \bibliography{oi_forest_paper}

\begin{thebibliography}{99}
\expandafter\ifx\csname natexlab\endcsname\relax\def\natexlab#1{#1}\fi

\bibitem[{{Abel} {et~al}\mbox{.}(1997){Abel}, {Anninos}, {Zhang}, \&
  {Norman}}]{abel1997}
{Abel} T., {Anninos} P., {Zhang} Y., {Norman} M.~L., 1997, \na, 2, 181

\bibitem[{{Akerman} {et~al}\mbox{.}(2005){Akerman}, {Ellison}, {Pettini}, \&
  {Steidel}}]{akerman2005}
{Akerman} C.~J., {Ellison} S.~L., {Pettini} M., {Steidel} C.~C., 2005, \aap,
  440, 499

\bibitem[{{Asplund} {et~al}\mbox{.}(2009){Asplund}, {Grevesse}, {Sauval}, \&
  {Scott}}]{asplund2009}
{Asplund} M., {Grevesse} N., {Sauval} A.~J., {Scott} P., 2009, \araa, 47, 481

\bibitem[{{Bahcall} \& {Peebles}(1969)}]{bahcall1969}
{Bahcall} J.~N., {Peebles} P.~J.~E., 1969, \apjl, 156, L7

\bibitem[{{Becker} \& {Bolton}(2013)}]{becker2013}
{Becker} G.~D., {Bolton} J.~S., 2013, \mnras, 436, 1023

\bibitem[{{Becker}, {Rauch} \& {Sargent}(2007){Becker}, {Rauch}, \&
  {Sargent}}]{becker2007}
{Becker} G.~D., {Rauch} M., {Sargent} W.~L.~W., 2007, \apj, 662, 72

\bibitem[{{Becker} {et~al}\mbox{.}(2011){Becker}, {Sargent}, {Rauch}, \&
  {Calverley}}]{becker2011}
{Becker} G.~D., {Sargent} W.~L.~W., {Rauch} M., {Calverley} A.~P., 2011, \apj,
  735, 93

\bibitem[{{Boisse} {et~al}\mbox{.}(1998){Boisse}, {Le Brun}, {Bergeron}, \&
  {Deharveng}}]{boisse1998}
{Boisse} P., {Le Brun} V., {Bergeron} J., {Deharveng} J.-M., 1998, \aap, 333,
  841

\bibitem[{{Bolton} \& {Becker}(2009)}]{bolton2009}
{Bolton} J.~S., {Becker} G.~D., 2009, \mnras, 398, L26

\bibitem[{{Bolton} \& {Haehnelt}(2007)}]{bolton2007}
{Bolton} J.~S., {Haehnelt} M.~G., 2007, \mnras, 374, 493

\bibitem[{{Bolton} \& {Haehnelt}(2013)}]{bolton2013}
{Bolton} J.~S., {Haehnelt} M.~G., 2013, \mnras, 429, 1695

\bibitem[{{Calverley} {et~al}\mbox{.}(2011){Calverley}, {Becker}, {Haehnelt},
  \& {Bolton}}]{calverley2011}
{Calverley} A.~P., {Becker} G.~D., {Haehnelt} M.~G., {Bolton} J.~S., 2011,
  \mnras, 412, 2543

\bibitem[{{Centuri{\'o}n} {et~al}\mbox{.}(2003){Centuri{\'o}n}, {Molaro},
  {Vladilo}, {P{\'e}roux}, {Levshakov}, \& {D'Odorico}}]{centurion2003}
{Centuri{\'o}n} M., {Molaro} P., {Vladilo} G., {P{\'e}roux} C., {Levshakov}
  S.~A., {D'Odorico} V., 2003, \aap, 403, 55

\bibitem[{{Churchill} {et~al}\mbox{.}(2000){Churchill}, {Mellon}, {Charlton},
  {Jannuzi}, {Kirhakos}, {Steidel}, \& {Schneider}}]{churchill2000}
{Churchill} C.~W., {Mellon} R.~R., {Charlton} J.~C., {Jannuzi} B.~T.,
  {Kirhakos} S., {Steidel} C.~C., {Schneider} D.~P., 2000, \apjs, 130, 91

\bibitem[{{Dessauges-Zavadsky} {et~al}\mbox{.}(2004){Dessauges-Zavadsky},
  {Calura}, {Prochaska}, {D'Odorico}, \& {Matteucci}}]{dessauges2004}
{Dessauges-Zavadsky} M., {Calura} F., {Prochaska} J.~X., {D'Odorico} S.,
  {Matteucci} F., 2004, \aap, 416, 79

\bibitem[{{Dessauges-Zavadsky} {et~al}\mbox{.}(2007){Dessauges-Zavadsky},
  {Calura}, {Prochaska}, {D'Odorico}, \& {Matteucci}}]{dessauges2007}
{Dessauges-Zavadsky} M., {Calura} F., {Prochaska} J.~X., {D'Odorico} S.,
  {Matteucci} F., 2007, \aap, 470, 431

\bibitem[{{Dessauges-Zavadsky} {et~al}\mbox{.}(2001){Dessauges-Zavadsky},
  {D'Odorico}, {McMahon}, {Molaro}, {Ledoux}, {P{\'e}roux}, \&
  {Storrie-Lombardi}}]{dessauges2001}
{Dessauges-Zavadsky} M., {D'Odorico} S., {McMahon} R.~G., {Molaro} P., {Ledoux}
  C., {P{\'e}roux} C., {Storrie-Lombardi} L.~J., 2001, \aap, 370, 426

\bibitem[{{Dessauges-Zavadsky} {et~al}\mbox{.}(2006){Dessauges-Zavadsky},
  {Prochaska}, {D'Odorico}, {Calura}, \& {Matteucci}}]{dessauges2006}
{Dessauges-Zavadsky} M., {Prochaska} J.~X., {D'Odorico} S., {Calura} F.,
  {Matteucci} F., 2006, \aap, 445, 93

\bibitem[{{Ellison} {et~al}\mbox{.}(2007){Ellison}, {Hennawi}, {Martin}, \&
  {Sommer-Larsen}}]{ellison2007}
{Ellison} S.~L., {Hennawi} J.~F., {Martin} C.~L., {Sommer-Larsen} J., 2007,
  \mnras, 378, 801

\bibitem[{{Ellison} {et~al}\mbox{.}(2001){Ellison}, {Pettini}, {Steidel}, \&
  {Shapley}}]{ellison2001}
{Ellison} S.~L., {Pettini} M., {Steidel} C.~C., {Shapley} A.~E., 2001, \apj,
  549, 770

\bibitem[{{Fan} {et~al}\mbox{.}(2006){Fan}, {Strauss}, {Becker}, {White},
  {Gunn}, {Knapp}, {Richards}, {Schneider}, {Brinkmann}, \&
  {Fukugita}}]{fan2006}
{Fan} X. {et~al.}, 2006, \aj, 132, 117

\bibitem[{{Ferland} {et~al}\mbox{.}(1998){Ferland}, {Korista}, {Verner},
  {Ferguson}, {Kingdon}, \& {Verner}}]{ferland1998}
{Ferland} G.~J., {Korista} K.~T., {Verner} D.~A., {Ferguson} J.~W., {Kingdon}
  J.~B., {Verner} E.~M., 1998, \pasp, 110, 761

\bibitem[{{Finlator} {et~al}\mbox{.}(2013){Finlator}, {Mu{\~n}oz},
  {Oppenheimer}, {Oh}, {{\"O}zel}, \& {Dav{\'e}}}]{finlator2013}
{Finlator} K., {Mu{\~n}oz} J.~A., {Oppenheimer} B.~D., {Oh} S.~P., {{\"O}zel}
  F., {Dav{\'e}} R., 2013, \mnras, 436, 1818

\bibitem[{{Fumagalli} {et~al}\mbox{.}(2013){Fumagalli}, {O'Meara}, {Prochaska},
  \& {Worseck}}]{fumagalli2013}
{Fumagalli} M., {O'Meara} J.~M., {Prochaska} J.~X., {Worseck} G., 2013, \apj,
  775, 78

\bibitem[{{Green}(2011)}]{green2011}
{Green} D.~A., 2011, Bulletin of the Astronomical Society of India, 39, 289

\bibitem[{{Haardt} \& {Madau}(2001)}]{haardt2001}
{Haardt} F., {Madau} P., 2001, in Clusters of Galaxies and the High Redshift
  Universe, {Neumann} D.~M., {Tran} J.~T.~V., eds.

\bibitem[{{Haardt} \& {Madau}(2012)}]{haardt2012}
{Haardt} F., {Madau} P., 2012, \apj, 746, 125

\bibitem[{{Haehnelt}, {Steinmetz} \& {Rauch}(1998){Haehnelt}, {Steinmetz}, \&
  {Rauch}}]{haehnelt1998}
{Haehnelt} M.~G., {Steinmetz} M., {Rauch} M., 1998, \apj, 495, 647

\bibitem[{{Jorgenson}, {Wolfe} \& {Prochaska}(2010){Jorgenson}, {Wolfe}, \&
  {Prochaska}}]{jorgenson2010}
{Jorgenson} R.~A., {Wolfe} A.~M., {Prochaska} J.~X., 2010, \apj, 722, 460

\bibitem[{{Khare} {et~al}\mbox{.}(2004){Khare}, {Kulkarni}, {Lauroesch},
  {York}, {Crotts}, \& {Nakamura}}]{khare2004}
{Khare} P., {Kulkarni} V.~P., {Lauroesch} J.~T., {York} D.~G., {Crotts}
  A.~P.~S., {Nakamura} O., 2004, \apj, 616, 86

\bibitem[{{Kohler} \& {Gnedin}(2007)}]{kohler2007}
{Kohler} K., {Gnedin} N.~Y., 2007, \apj, 655, 685

\bibitem[{{Kulkarni} {et~al}\mbox{.}(2013){Kulkarni}, {Rollinde}, {Hennawi}, \&
  {Vangioni}}]{kulkarni2013}
{Kulkarni} G., {Rollinde} E., {Hennawi} J.~F., {Vangioni} E., 2013, \apj, 772,
  93

\bibitem[{{Kulkarni} {et~al}\mbox{.}(2005){Kulkarni}, {Fall}, {Lauroesch},
  {York}, {Welty}, {Khare}, \& {Truran}}]{kulkarni2005}
{Kulkarni} V.~P., {Fall} S.~M., {Lauroesch} J.~T., {York} D.~G., {Welty} D.~E.,
  {Khare} P., {Truran} J.~W., 2005, \apj, 618, 68

\bibitem[{{Ledoux}, {Bergeron} \& {Petitjean}(2002){Ledoux}, {Bergeron}, \&
  {Petitjean}}]{ledoux2002a}
{Ledoux} C., {Bergeron} J., {Petitjean} P., 2002, \aap, 385, 802

\bibitem[{{Ledoux} {et~al}\mbox{.}(2006){Ledoux}, {Petitjean}, {Fynbo},
  {M{\o}ller}, \& {Srianand}}]{ledoux2006}
{Ledoux} C., {Petitjean} P., {Fynbo} J.~P.~U., {M{\o}ller} P., {Srianand} R.,
  2006, \aap, 457, 71

\bibitem[{{Ledoux}, {Petitjean} \& {Srianand}(2003){Ledoux}, {Petitjean}, \&
  {Srianand}}]{ledoux2003}
{Ledoux} C., {Petitjean} P., {Srianand} R., 2003, \mnras, 346, 209

\bibitem[{{Ledoux}, {Srianand} \& {Petitjean}(2002){Ledoux}, {Srianand}, \&
  {Petitjean}}]{ledoux2002b}
{Ledoux} C., {Srianand} R., {Petitjean} P., 2002, \aap, 392, 781

\bibitem[{{Levshakov} {et~al}\mbox{.}(2002){Levshakov}, {Dessauges-Zavadsky},
  {D'Odorico}, \& {Molaro}}]{levshakov2002}
{Levshakov} S.~A., {Dessauges-Zavadsky} M., {D'Odorico} S., {Molaro} P., 2002,
  \apj, 565, 696

\bibitem[{{Lopez} \& {Ellison}(2003)}]{lopez2003}
{Lopez} S., {Ellison} S.~L., 2003, \aap, 403, 573

\bibitem[{{Lopez} {et~al}\mbox{.}(2002){Lopez}, {Reimers}, {D'Odorico}, \&
  {Prochaska}}]{lopez2002}
{Lopez} S., {Reimers} D., {D'Odorico} S., {Prochaska} J.~X., 2002, \aap, 385,
  778

\bibitem[{{Lopez} {et~al}\mbox{.}(1999){Lopez}, {Reimers}, {Rauch}, {Sargent},
  \& {Smette}}]{lopez1999}
{Lopez} S., {Reimers} D., {Rauch} M., {Sargent} W.~L.~W., {Smette} A., 1999,
  \apj, 513, 598

\bibitem[{{Lu}, {Sargent} \& {Barlow}(1998){Lu}, {Sargent}, \&
  {Barlow}}]{lu1998}
{Lu} L., {Sargent} W.~L.~W., {Barlow} T.~A., 1998, \aj, 115, 55

\bibitem[{{Lu} {et~al}\mbox{.}(1996){Lu}, {Sargent}, {Barlow}, {Churchill}, \&
  {Vogt}}]{lu1996}
{Lu} L., {Sargent} W.~L.~W., {Barlow} T.~A., {Churchill} C.~W., {Vogt} S.~S.,
  1996, \apjs, 107, 475

\bibitem[{{Maio}, {Ciardi} \& {M{\"u}ller}(2013){Maio}, {Ciardi}, \&
  {M{\"u}ller}}]{maio2013}
{Maio} U., {Ciardi} B., {M{\"u}ller} V., 2013, \mnras, 435, 1443

\bibitem[{{McQuinn} {et~al}\mbox{.}(2008){McQuinn}, {Lidz}, {Zaldarriaga},
  {Hernquist}, \& {Dutta}}]{mcquinn2008}
{McQuinn} M., {Lidz} A., {Zaldarriaga} M., {Hernquist} L., {Dutta} S., 2008,
  \mnras, 388, 1101

\bibitem[{{McQuinn}, {Oh} \& {Faucher-Gigu{\`e}re}(2011){McQuinn}, {Oh}, \&
  {Faucher-Gigu{\`e}re}}]{mcquinn2011}
{McQuinn} M., {Oh} S.~P., {Faucher-Gigu{\`e}re} C.-A., 2011, \apj, 743, 82

\bibitem[{{Meiring} {et~al}\mbox{.}(2006){Meiring}, {Kulkarni}, {Khare},
  {Bechtold}, {York}, {Cui}, {Lauroesch}, {Crotts}, \&
  {Nakamura}}]{meiring2006}
{Meiring} J.~D. {et~al.}, 2006, \mnras, 370, 43

\bibitem[{{Meiring} {et~al}\mbox{.}(2007){Meiring}, {Lauroesch}, {Kulkarni},
  {P{\'e}roux}, {Khare}, {York}, \& {Crotts}}]{meiring2007}
{Meiring} J.~D., {Lauroesch} J.~T., {Kulkarni} V.~P., {P{\'e}roux} C., {Khare}
  P., {York} D.~G., {Crotts} A.~P.~S., 2007, \mnras, 376, 557

\bibitem[{{Meiring} {et~al}\mbox{.}(2011){Meiring}, {Tripp}, {Prochaska},
  {Tumlinson}, {Werk}, {Jenkins}, {Thom}, {O'Meara}, \&
  {Sembach}}]{meiring2011}
{Meiring} J.~D. {et~al.}, 2011, \apj, 732, 35

\bibitem[{{Mesinger}(2010)}]{mesinger2010}
{Mesinger} A., 2010, \mnras, 407, 1328

\bibitem[{{Meyer}, {Lanzetta} \& {Wolfe}(1995){Meyer}, {Lanzetta}, \&
  {Wolfe}}]{meyer1995}
{Meyer} D.~M., {Lanzetta} K.~M., {Wolfe} A.~M., 1995, \apjl, 451, L13

\bibitem[{{Miralda-Escud{\'e}}, {Haehnelt} \& {Rees}(2000){Miralda-Escud{\'e}},
  {Haehnelt}, \& {Rees}}]{miralda2000}
{Miralda-Escud{\'e}} J., {Haehnelt} M., {Rees} M.~J., 2000, \apj, 530, 1

\bibitem[{{Molaro} {et~al}\mbox{.}(2000){Molaro}, {Bonifacio}, {Centuri{\'o}n},
  {D'Odorico}, {Vladilo}, {Santin}, \& {Di Marcantonio}}]{molaro2000}
{Molaro} P., {Bonifacio} P., {Centuri{\'o}n} M., {D'Odorico} S., {Vladilo} G.,
  {Santin} P., {Di Marcantonio} P., 2000, \apj, 541, 54

\bibitem[{{Molaro} {et~al}\mbox{.}(2001){Molaro}, {Levshakov}, {D'Odorico},
  {Bonifacio}, \& {Centuri{\'o}n}}]{molaro2001}
{Molaro} P., {Levshakov} S.~A., {D'Odorico} S., {Bonifacio} P., {Centuri{\'o}n}
  M., 2001, \apj, 549, 90

\bibitem[{{Mortlock} {et~al}\mbox{.}(2011){Mortlock}, {Warren}, {Venemans},
  {Patel}, {Hewett}, {McMahon}, {Simpson}, {Theuns}, {Gonz{\'a}les-Solares},
  {Adamson}, {Dye}, {Hambly}, {Hirst}, {Irwin}, {Kuiper}, {Lawrence}, \&
  {R{\"o}ttgering}}]{mortlock2011}
{Mortlock} D.~J. {et~al.}, 2011, \nat, 474, 616

\bibitem[{{Nestor} {et~al}\mbox{.}(2008){Nestor}, {Pettini}, {Hewett}, {Rao},
  \& {Wild}}]{nestor2008}
{Nestor} D.~B., {Pettini} M., {Hewett} P.~C., {Rao} S., {Wild} V., 2008,
  \mnras, 390, 1670

\bibitem[{{Noterdaeme} {et~al}\mbox{.}(2008){Noterdaeme}, {Ledoux},
  {Petitjean}, \& {Srianand}}]{noterdaeme2008}
{Noterdaeme} P., {Ledoux} C., {Petitjean} P., {Srianand} R., 2008, \aap, 481,
  327

\bibitem[{{Noterdaeme} {et~al}\mbox{.}(2012){Noterdaeme}, {Petitjean},
  {Carithers}, {P{\^a}ris}, {Font-Ribera}, {Bailey}, {Aubourg}, {Bizyaev},
  {Ebelke}, {Finley}, {Ge}, {Malanushenko}, {Malanushenko},
  {Miralda-Escud{\'e}}, {Myers}, {Oravetz}, {Pan}, {Pieri}, {Ross},
  {Schneider}, {Simmons}, \& {York}}]{noterdaeme2012}
{Noterdaeme} P. {et~al.}, 2012, \aap, 547, L1

\bibitem[{{Oh}(2002)}]{oh2002}
{Oh} S.~P., 2002, \mnras, 336, 1021

\bibitem[{{O'Meara} {et~al}\mbox{.}(2013){O'Meara}, {Prochaska}, {Worseck},
  {Chen}, \& {Madau}}]{omeara2013}
{O'Meara} J.~M., {Prochaska} J.~X., {Worseck} G., {Chen} H.-W., {Madau} P.,
  2013, \apj, 765, 137

\bibitem[{{P{\'e}roux} {et~al}\mbox{.}(2006){P{\'e}roux}, {Meiring},
  {Kulkarni}, {Ferlet}, {Khare}, {Lauroesch}, {Vladilo}, \&
  {York}}]{peroux2006}
{P{\'e}roux} C., {Meiring} J.~D., {Kulkarni} V.~P., {Ferlet} R., {Khare} P.,
  {Lauroesch} J.~T., {Vladilo} G., {York} D.~G., 2006, \mnras, 372, 369

\bibitem[{{P{\'e}roux} {et~al}\mbox{.}(2008){P{\'e}roux}, {Meiring},
  {Kulkarni}, {Khare}, {Lauroesch}, {Vladilo}, \& {York}}]{peroux2008}
{P{\'e}roux} C., {Meiring} J.~D., {Kulkarni} V.~P., {Khare} P., {Lauroesch}
  J.~T., {Vladilo} G., {York} D.~G., 2008, \mnras, 386, 2209

\bibitem[{{Petitjean}, {Srianand} \& {Ledoux}(2000){Petitjean}, {Srianand}, \&
  {Ledoux}}]{petitjean2000}
{Petitjean} P., {Srianand} R., {Ledoux} C., 2000, \aap, 364, L26

\bibitem[{{Petitjean}, {Srianand} \& {Ledoux}(2002){Petitjean}, {Srianand}, \&
  {Ledoux}}]{petitjean2002}
{Petitjean} P., {Srianand} R., {Ledoux} C., 2002, \mnras, 332, 383

\bibitem[{{Pettini} {et~al}\mbox{.}(1999){Pettini}, {Ellison}, {Steidel}, \&
  {Bowen}}]{pettini1999}
{Pettini} M., {Ellison} S.~L., {Steidel} C.~C., {Bowen} D.~V., 1999, \apj, 510,
  576

\bibitem[{{Pettini} {et~al}\mbox{.}(2000){Pettini}, {Ellison}, {Steidel},
  {Shapley}, \& {Bowen}}]{pettini2000}
{Pettini} M., {Ellison} S.~L., {Steidel} C.~C., {Shapley} A.~E., {Bowen} D.~V.,
  2000, \apj, 532, 65

\bibitem[{{Prochaska} \& {Burles}(1999)}]{prochaska1999b}
{Prochaska} J.~X., {Burles} S.~M., 1999, \aj, 117, 1957

\bibitem[{{Prochaska}, {Gawiser} \& {Wolfe}(2001){Prochaska}, {Gawiser}, \&
  {Wolfe}}]{prochaska2001a}
{Prochaska} J.~X., {Gawiser} E., {Wolfe} A.~M., 2001, \apj, 552, 99

\bibitem[{{Prochaska} {et~al}\mbox{.}(2003{\natexlab{a}}){Prochaska},
  {Gawiser}, {Wolfe}, {Castro}, \& {Djorgovski}}]{prochaska2003a}
{Prochaska} J.~X., {Gawiser} E., {Wolfe} A.~M., {Castro} S., {Djorgovski}
  S.~G., 2003{\natexlab{a}}, \apjl, 595, L9

\bibitem[{{Prochaska} {et~al}\mbox{.}(2003{\natexlab{b}}){Prochaska},
  {Gawiser}, {Wolfe}, {Cooke}, \& {Gelino}}]{prochaska2003b}
{Prochaska} J.~X., {Gawiser} E., {Wolfe} A.~M., {Cooke} J., {Gelino} D.,
  2003{\natexlab{b}}, \apjs, 147, 227

\bibitem[{{Prochaska} \& {Wolfe}(1996)}]{prochaska1996}
{Prochaska} J.~X., {Wolfe} A.~M., 1996, \apj, 470, 403

\bibitem[{{Prochaska} \& {Wolfe}(1997)}]{prochaska1997}
{Prochaska} J.~X., {Wolfe} A.~M., 1997, \apj, 474, 140

\bibitem[{{Prochaska} \& {Wolfe}(2000)}]{prochaska2000}
{Prochaska} J.~X., {Wolfe} A.~M., 2000, \apjl, 533, L5

\bibitem[{{Prochaska} \& {Wolfe}(2002)}]{prochaska2002}
{Prochaska} J.~X., {Wolfe} A.~M., 2002, \apj, 566, 68

\bibitem[{{Prochaska} \& {Wolfe}(2009)}]{prochaska2009}
{Prochaska} J.~X., {Wolfe} A.~M., 2009, \apj, 696, 1543

\bibitem[{{Prochaska} {et~al}\mbox{.}(2007){Prochaska}, {Wolfe}, {Howk},
  {Gawiser}, {Burles}, \& {Cooke}}]{prochaska2007}
{Prochaska} J.~X., {Wolfe} A.~M., {Howk} J.~C., {Gawiser} E., {Burles} S.~M.,
  {Cooke} J., 2007, \apjs, 171, 29

\bibitem[{{Prochaska} {et~al}\mbox{.}(2001){Prochaska}, {Wolfe}, {Tytler},
  {Burles}, {Cooke}, {Gawiser}, {Kirkman}, {O'Meara}, \&
  {Storrie-Lombardi}}]{prochaska2001b}
{Prochaska} J.~X. {et~al.}, 2001, \apjs, 137, 21

\bibitem[{{Rafelski} {et~al}\mbox{.}(2012){Rafelski}, {Wolfe}, {Prochaska},
  {Neeleman}, \& {Mendez}}]{rafelski2012}
{Rafelski} M., {Wolfe} A.~M., {Prochaska} J.~X., {Neeleman} M., {Mendez} A.~J.,
  2012, \apj, 755, 89

\bibitem[{{Rahmati} {et~al}\mbox{.}(2013{\natexlab{a}}){Rahmati}, {Pawlik},
  {Rai{\v c}evi\`{c}}, \& {Schaye}}]{rahmati2013}
{Rahmati} A., {Pawlik} A.~H., {Rai{\v c}evi\`{c}} M., {Schaye} J.,
  2013{\natexlab{a}}, \mnras, 430, 2427

\bibitem[{{Rahmati} {et~al}\mbox{.}(2013{\natexlab{b}}){Rahmati}, {Schaye},
  {Pawlik}, \& {Rai{\v c}evi\`{c}}}]{rahmati2013b}
{Rahmati} A., {Schaye} J., {Pawlik} A.~H., {Rai{\v c}evi\`{c}} M.,
  2013{\natexlab{b}}, \mnras, 431, 2261

\bibitem[{{Rao} {et~al}\mbox{.}(2005){Rao}, {Prochaska}, {Howk}, \&
  {Wolfe}}]{rao2005}
{Rao} S.~M., {Prochaska} J.~X., {Howk} J.~C., {Wolfe} A.~M., 2005, \aj, 129, 9

\bibitem[{{Rao} \& {Turnshek}(2000)}]{rao2000}
{Rao} S.~M., {Turnshek} D.~A., 2000, \apjs, 130, 1

\bibitem[{{Rao}, {Turnshek} \& {Nestor}(2006){Rao}, {Turnshek}, \&
  {Nestor}}]{rao2006}
{Rao} S.~M., {Turnshek} D.~A., {Nestor} D.~B., 2006, \apj, 636, 610

\bibitem[{{Ribaudo}, {Lehner} \& {Howk}(2011){Ribaudo}, {Lehner}, \&
  {Howk}}]{ribaudo2011}
{Ribaudo} J., {Lehner} N., {Howk} J.~C., 2011, \apj, 736, 42

\bibitem[{{Schaye}(2001)}]{schaye2001}
{Schaye} J., 2001, \apj, 559, 507

\bibitem[{{Schaye} {et~al}\mbox{.}(2003){Schaye}, {Aguirre}, {Kim}, {Theuns},
  {Rauch}, \& {Sargent}}]{schaye2003}
{Schaye} J., {Aguirre} A., {Kim} T.-S., {Theuns} T., {Rauch} M., {Sargent}
  W.~L.~W., 2003, \apj, 596, 768

\bibitem[{{Seyffert} {et~al}\mbox{.}(2013){Seyffert}, {Cooksey}, {Simcoe},
  {O'Meara}, {Kao}, \& {Prochaska}}]{seyffert2013}
{Seyffert} E.~N., {Cooksey} K.~L., {Simcoe} R.~A., {O'Meara} J.~M., {Kao}
  M.~M., {Prochaska} J.~X., 2013, ArXiv e-prints

\bibitem[{{Simcoe}(2011)}]{simcoe2011}
{Simcoe} R.~A., 2011, \apj, 738, 159

\bibitem[{{Simcoe} {et~al}\mbox{.}(2012){Simcoe}, {Sullivan}, {Cooksey}, {Kao},
  {Matejek}, \& {Burgasser}}]{simcoe2012}
{Simcoe} R.~A., {Sullivan} P.~W., {Cooksey} K.~L., {Kao} M.~M., {Matejek}
  M.~S., {Burgasser} A.~J., 2012, \nat, 492, 79

\bibitem[{{Songaila}(2004)}]{songaila2004}
{Songaila} A., 2004, \aj, 127, 2598

\bibitem[{{Songaila} \& {Cowie}(2002)}]{songaila2002}
{Songaila} A., {Cowie} L.~L., 2002, \aj, 123, 2183

\bibitem[{{Songaila} \& {Cowie}(2010)}]{songaila2010}
{Songaila} A., {Cowie} L.~L., 2010, \apj, 721, 1448

\bibitem[{{Springel}(2005)}]{springel2005}
{Springel} V., 2005, \mnras, 364, 1105

\bibitem[{{Srianand}, {Petitjean} \& {Ledoux}(2000){Srianand}, {Petitjean}, \&
  {Ledoux}}]{srianand2000}
{Srianand} R., {Petitjean} P., {Ledoux} C., 2000, \nat, 408, 931

\bibitem[{{Turnshek} {et~al}\mbox{.}(2004){Turnshek}, {Rao}, {Nestor}, {Vanden
  Berk}, {Belfort-Mihalyi}, \& {Monier}}]{turnshek2004}
{Turnshek} D.~A., {Rao} S.~M., {Nestor} D.~B., {Vanden Berk} D.,
  {Belfort-Mihalyi} M., {Monier} E.~M., 2004, \apjl, 609, L53

\bibitem[{{Vladilo} {et~al}\mbox{.}(2011){Vladilo}, {Abate}, {Yin}, {Cescutti},
  \& {Matteucci}}]{vladilo2011}
{Vladilo} G., {Abate} C., {Yin} J., {Cescutti} G., {Matteucci} F., 2011, \aap,
  530, A33

\bibitem[{{Wolfe} {et~al}\mbox{.}(1994){Wolfe}, {Fan}, {Tytler}, {Vogt},
  {Keane}, \& {Lanzetta}}]{wolfe1994}
{Wolfe} A.~M., {Fan} X.-M., {Tytler} D., {Vogt} S.~S., {Keane} M.~J.,
  {Lanzetta} K.~M., 1994, \apjl, 435, L101

\bibitem[{{Wolfe} {et~al}\mbox{.}(2008){Wolfe}, {Prochaska}, {Jorgenson}, \&
  {Rafelski}}]{wolfe2008}
{Wolfe} A.~M., {Prochaska} J.~X., {Jorgenson} R.~A., {Rafelski} M., 2008, \apj,
  681, 881

\bibitem[{{Wyithe} \& {Bolton}(2011)}]{wyithe2011}
{Wyithe} J.~S.~B., {Bolton} J.~S., 2011, \mnras, 412, 1926

\end{thebibliography}

\bsp

\appendix

\section{O\,{\sevensize\bf I} as a tracer of H\,{\sevensize\bf I}}

We have used the photo-ionization code CLOUDY \citep{ferland1998} to
estimate the  neutral oxygen fraction as a function of  \hi column
density. The blue solid curve shows the the ratio of the neutral
fractions of oxygen and hydrogen   $f_{\rmn{OI}/\rmn{HI}}= f_{\rmn{OI}}/f_{\rmn{HI}}$
for a slab of constant density illuminated from the
outside  with  the \citet{haardt2012} model for the UV background, 
which includes contributions from quasars and galaxies. The simulations were performed at $z = 6$ and the effects of
the CMB and  background cosmic rays were also taken into account. 
We further  assumed an equation of state with constant pressure.
The result is very similar to that obtained  for the 
\citet{haardt2001} UV background model shown by the black curve.  In order to test the
sensitivity to the amplitude of of the UV background and the
temperature of the gas we also show the results with the UV 
intensity increased/decreased by a factor of three and for fixed
temperatures of 5000K and 10000K  with the dot-dashed and dotted
curves  as indicated on the plot. The results were found to be insensitive to the number density of hydrogen in the slab.

For column densities $\log N_{\rmn{HI}} \ga 17$, where the gas is self-shielded, 
the neutral fraction of oxygen  traces that of hydrogen very well 
(to within 0.1 dex or better) independent of the detailed  assumptions
of our CLOUDY calculations. This is unsurprising since at $z \sim 6$, the main contributuion to the ionizing background is from galaxies and is therefore relatively soft. For harder ionizing spectra, $f_{\rmn{OI}/\rmn{HI}}$ would begin to decline at higher column densities. At smaller \hi column densities,  the neutral
fraction  of oxygen drops significantly faster than that of hydrogen 
and  $f_{\rmn{OI}/\rmn{HI}}$ depends  on the amplitude of the UV
background, but is very similar for the two background UV models we
tested.  It  also changes little for our two constant temperature calculations. We should note here  that non-selfshielded regions
contribute, however,   negligibly to the \oi absorbers discussed in the main paper. 

We have adopted the calculation with the  \citet{haardt2012} model of
the UV background as our fiducial model for the calculation of the 
\oi absorption in our simulations.  We produced a set of interpolation
tables that were used in the relationship   $n_{\rmn{OI}} = Z_{\rmn O}
\, f_{\rmn{OI/HI}} \, n_{\rmn{HI}}$ (Section 3.1)  for each \oi absorber.

\begin{figure}
\includegraphics[width=1.0\columnwidth]{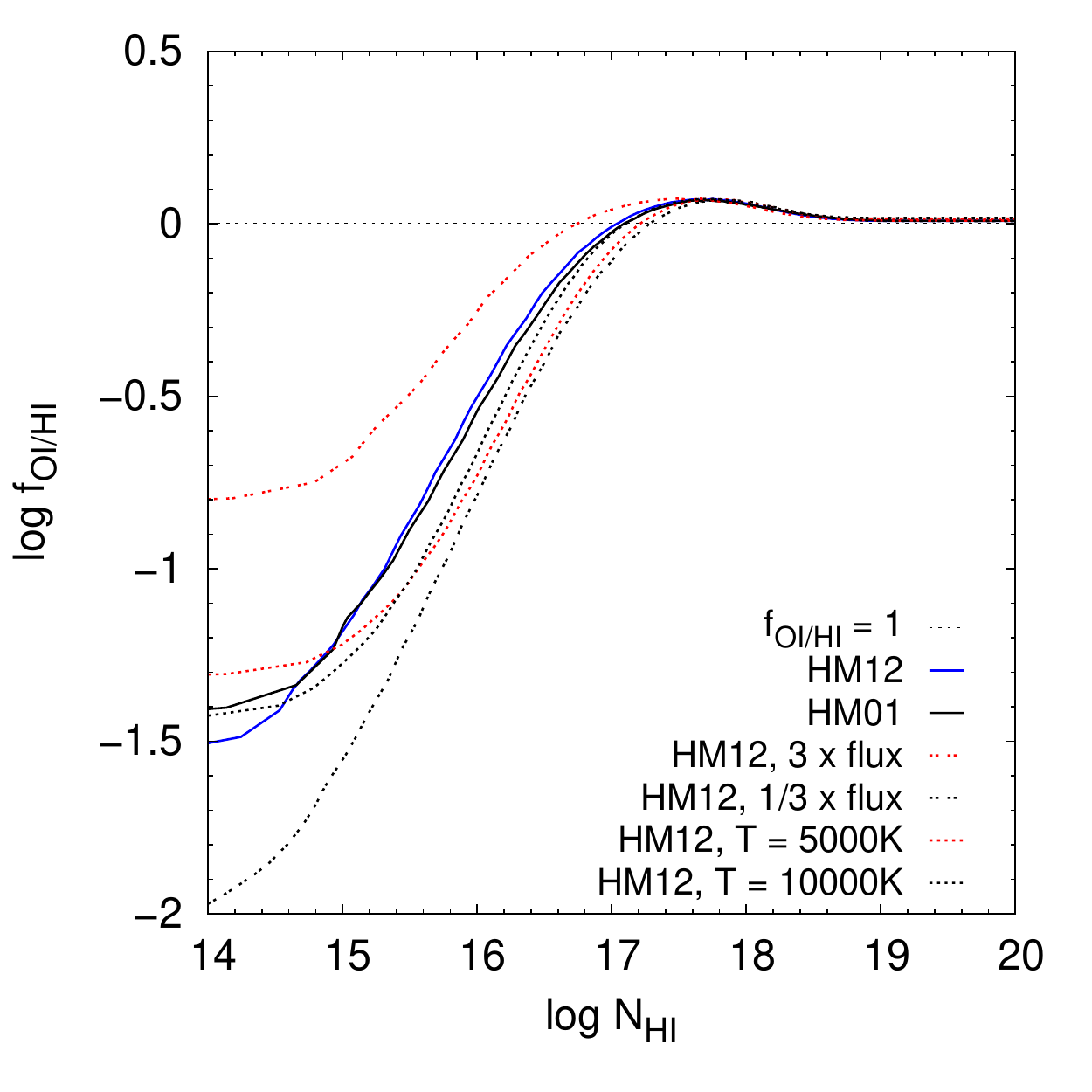}
\caption{Plot of the neutral fraction of oxygen relative to the
  neutral fraction of hydrogen against \hi column density. The blue
  curve  assumes the \citet{haardt2012} UV backround model. We find that \oi
  is a good tracer of \hi for self-shielded gas with $\log
  N_{\rmn{HI}} \ga 17$. Calculations for the \citet{haardt2001} model
  of the UV  background and for decreased/increased amplitude of the
  UV background and constant temperature of the gas of 5000K and 10000K 
  are also shown. The dashed line  shows $f_{\rmn{OI}/\rmn{HI}} = 1$.}
\label{oifrac}
\end{figure}

\label{lastpage}

\end{document}